\documentclass[a4paper,german,structabstract]{aa}

\usepackage{natbib}
\bibpunct{(}{)}{;}{a}{}{,}

\usepackage{multicol}
\usepackage{amsmath}
\usepackage{amssymb}
\usepackage{wasysym}
\usepackage{epsfig}
\usepackage{epsf}
\usepackage[dvips]{color}
\usepackage{graphicx,graphics}
\usepackage{float}
\usepackage{setspace,lineno}
\usepackage[latin1]{inputenc}
\usepackage{txfonts}

\begin{document}
\begin{multicols}{1}

\title{Spectroscopic characterization of the atmospheres of potentially habitable planets: GL\,581\,d as a model case study}
\titlerunning{Characterizing potentially habitable planets}

\author{P. von Paris\inst{1,2,3} \and
J. Cabrera\inst{1} \and M. Godolt\inst{4} \and J.L. Grenfell\inst{4}
\and P. Hedelt\inst{2,3} \and H. Rauer\inst{1,4} \and F.
Schreier\inst{5} \and B. Stracke\inst{1}}

\institute{Institut f\"{u}r Planetenforschung, Deutsches Zentrum
f\"{u}r Luft- und Raumfahrt, Rutherfordstr. 2, 12489 Berlin, Germany
\and Laboratoire d'Astrophysique de Bordeaux, 2 rue de
l'Observatoire BP 89 - 33271 Floirac Cedex, France \and Universit\'e
de Bordeaux, Observatoire Aquitain des Sciences de l'Univers, 2 rue
de l'Observatoire BP 89 - 33271 Floirac Cedex, France 
\and Zentrum f\"{u}r Astronomie und Astrophysik, Technische
Universit\"{a}t Berlin, Hardenbergstr. 36, 10623 Berlin, Germany
\and Institut
f\"{u}r Methodik der Fernerkundung, Deutsches Zentrum f\"{u}r Luft-
und Raumfahrt,  Oberpfaffenhoffen, 82234 We{\ss}ling, Germany}

\abstract {Were a potentially habitable planet to be discovered, the
next step would be the search for an atmosphere and its
characterization. Eventually, surface conditions, hence
habitability, and biomarkers as indicators for life would be
assessed.} {The Super-Earth candidate Gliese (GL) 581\,d is the first
potentially habitable extrasolar planet discovered so far. Therefore, GL\,581\,d is used
to illustrate a hypothetical detailed spectroscopic characterization of such
planets.} {Atmospheric profiles from a wide range of possible 1D radiative-convective model
scenarios of GL\,581\,d were used to calculate high-resolution
synthetic emission and transmission spectra.  Atmospheres were
assumed to be composed of N$_2$, CO$_2$ and H$_2$O. From the spectra, signal-to-noise ratios (SNR) were calculated for a telescope such as the planned James Webb Space Telescope (JWST). Exposure times were set equal to the duration of one transit.}
 {The presence of the model atmospheres could be clearly inferred from the calculated synthetic spectra due to strong water and carbon dioxide absorption
bands. Surface temperatures could be inferred for model scenarios with optically thin spectral windows. Dense, CO$_2$-rich (potentially habitable) scenarios did not allow for the characterization of surface temperatures and to assess habitability. Degeneracies between CO$_2$ concentration and surface pressure further complicated the interpretation of the calculated spectra, hence the determination of atmospheric conditions. Still, inferring approximative CO$_2$ concentrations and surface pressures would be possible.

In practice, detecting atmospheric signals is challenging since calculated
SNR values are well below unity in most of the cases. The SNR for a single transit was
only barely larger than unity in some near-IR bands for transmission spectroscopy.

Most interestingly, the false-positive detection of biomarker
candidates such as methane and ozone could be possible in low
resolution spectra due to the presence of CO$_2$ absorption bands
which overlap with biomarker spectral bands. This can be avoided
however by observing all main CO$_2$ IR bands instead of
concentrating on, e.g., the 4.3 or 15$\mu$m bands only.  Furthermore, a masking of ozone signatures by CO$_2$ absorption bands is shown to be possible. Simulations imply that such a false-negative detection of ozone  would be possible even for rather large ozone concentrations of up to 10$^{-5}$.}{}


\maketitle

\end{multicols}{1}

\section{Introduction}

Currently, more than 500 extrasolar planets and planet candidates
are known. Over 100 of these planets transit their central star. Ten
of the transiting planets are so-called Super-Earths with masses
below 10 Earth masses: CoRoT-7 b \citep{leger2009}, GJ\,1214\,b
\citep{charb2009}, Kepler-9 d (\citealp{holman2010},
\citealp{torres2011}), Kepler-10 b \citep{batalha2011},
Kepler-10 c \citep{fressin2011}, Kepler-11 b, d, e and
f \citep{lissauer2011}  and 55 Cnc e (\citealp{winn2011},
\citealp{demory2011}).

The unique geometrical orientation of transiting planets offers the
opportunity for the spectral characterization of their atmospheres.
Transmission spectroscopy during the primary transit favors ultraviolet (UV),
visible and near-infrared (IR) wavelengths, since the stellar signal is
stronger towards shorter wavelengths. Emission spectroscopy during
the secondary eclipse is easier in the mid-IR since the planet-star
flux ratio is higher in this wavelength regime. Thus, both methods
are complementary.

Emission and transmission observations have been performed for
almost 30 extrasolar planets to date. Thermal emission of radiation
has been detected (e.g., \citealp{Deming2005},
\citealp{harrington2006}, \citealp{knutson2007daynight_189733},
\citealp{alonso2009}). Furthermore, the chemical composition of the
atmospheres and exospheres of some of these planets has been
determined. Atoms (H, C, O, Na, Fe, Mg) and molecules (CO, CO$_2$,
CH$_4$, H$_2$O) have been detected (e.g., \citealp{charb2002},
\citealp{vidal2004}, \citealp{tinetti2007}, \citealp{grillmair2008},
\citealp{swain2009}, \citealp{stevenson2010},
\citealp{madhusudhan2011}). Although there is some discussion
on-going as to whether some of these observations are valid
(\citealp{gibson2011}, \citealp{mandell2011}), atmospheric
characterization of exoplanets is indeed feasible with current
instrumentation. For two transiting Super-Earths, spectroscopic
observations have been already performed (CoRoT-7 b,
\citealp{guenther2011}, and GJ\,1214\,b, \citealp{bean2010},
\citealp{desert2011}, \citealp{croll2011gj1214}, \citealp{crossfield2011}). For CoRoT-7 b,
upper limits on the extension of the exosphere have been obtained.
The observations of GJ\,1214\,b narrowed down the range of possible
atmospheric scenarios, favoring either a cloud-free water vapor
atmosphere or a cloudy hydrogen-dominated, methane-depleted atmosphere. These two
planets are however far too hot to be considered habitable in the
classical sense of life as we know it on Earth.

Nevertheless, studies of potential atmospheric signatures of
terrestrial habitable planets have been performed, in order to
predict signal strengths and assess observation strategies (e.g.,
\citealp{DesMarais2002}, \citealp{Seg2003},
\citealp{ehrenreich2006}, \citealp{Kaltenegger2009},
\citealp{miller_ricci2009_atmo}, \citealp{deming2009},
\citealp{belu2010}, \citealp{rauer2011}).\newline

The extrasolar planet GL\,581\,d (\citealp{udry2007},
\citealp{mayor2009gliese}) is the first potentially habitable
Super-Earth (\citealp{wordsworth2010}, \citealp{vparis2010gliese},
\citealp{hu2011}, \citealp{kaltenegger2011}, \citealp{wordsworth2011}). The orbital inclination of GL\,581\,d has been shown to lie in a range between 40-88$^{\circ}$,
based on photometric constraints for GL\,581\,b \citep{lopez2006} and
dynamical simulations of the whole system \citep{mayor2009gliese}.
Hence, GL\,581\,d is unlikely to transit its central star. However, we
used GL\,581\,d as an analogue of similar, transiting systems which
are anticipated to be found in the near future. In this study, we
illustrate the possible spectroscopic characterization of
potentially habitable planets based on a wide range of atmospheric model scenarios
of GL\,581\,d from \citet{vparis2010gliese}, following the
strategy outlined below. Synthetic spectra of some specific, potentially habitable GL\,581\,d model
atmospheres have already been presented by \citet{kaltenegger2011} and \citet{wordsworth2011}.
However, they did not discuss the potential detectability (i.e.,
signal-to-noise ratios) of spectral features or the potentially
possible detailed characterization of their model atmospheres.

The paper is organized as follows: Section \ref{observers} briefly outlines the mentioned observing strategy. Section \ref{theatmosphere}
presents atmospheric scenarios and models used. Results will be
shown in Sect. \ref{resultsect}. Conclusions are given in Sect.
\ref{concl}.

\section{Potential atmospheric characterization of terrestrial exoplanets}

\label{observers}

\subsection{Observations}

The observable quantity during transit measurements is the wavelength-dependent planet-to-star contrast ratio, i.e. the transit or eclipse depth. Based on the knowledge of the stellar radius, transmission spectra thus measure the apparent planetary radius as a function of wavelength. However, this depends critically on the accurate characterization of the central star, which is the main source of uncertainty for derived planetary properties (see, e.g., the case of GJ\,1214\,b, \citealp{carter2011}). From the stellar properties such as spectrum and radius, and adopting a baseline value for the planetary radius, emission spectra during secondary eclipse can be translated into brightness temperature spectra. The brightness temperature spectra are particularly illustrative since the brightness temperature is the apparent atmospheric temperature at a given wavelength. For optically thin spectral windows, the brightness temperature corresponds to the surface
temperature.\newline Note that in the following, we assume that the stellar properties as well as the geometric planet radius are known exactly.

\subsection{Atmospheric characterization}

An observation strategy would aim at (1) establishing the existence of an atmosphere, (2) determining the major atmospheric constituents along with radiative trace gases, (3) characterizing surface conditions, hence assessing habitability and ultimately, (4) searching for atmospheric species which would indicate the presence of life on a planet. Usually, however, the existence of an atmosphere and its composition can only be established via atmospheric and spectral modeling with subsequent comparison to the obtained data (e.g., \citealp{madhusudhan2009}, \citealp{miller_ricci2009_atmo}, \citealp{miller_ricci2010}).\newline An additional challenge is the possible false-positive or
false-negative identification of so-called biomarkers (e.g., O$_3$,
CH$_4$, N$_2$O) which was discussed, e.g., by \citet{selsis2002} or \citet{schindler2000}.
Biomarkers are atmospheric species assumed to be indicative of the
presence of a biosphere on the planet. On Earth, N$_2$O, for
example, is believed to be almost exclusively produced by
denitrifying bacteria, and O$_3$ is a photochemical product of
oxygen which itself originates mainly from photosynthetic organisms.
Such biomarkers are detectable in the Earth's atmosphere due to
absorption bands (ozone: 9.6 $\mu$m, nitrous oxide: 7.8 and 4.5
$\mu$m, methane: 7.7 and 3.3 $\mu$m). \newline It has to be distinguished between the false-positive or false-negative detection of biomarker species and the false-positive interpretation of detected biomarkers as a sign for life. This is due to the fact that abiotic formation of ozone is possible
(e.g., \citealp{segura2007}, \citealp{domagal2010}) even though the magnitude of the effect is
debated \citep{selsis2002}. Note, for example, that ozone has been found in the Martian atmosphere (e.g.
\citealp{yung1999}). Therefore, the detection of ozone alone could
still be a false-positive detection of life. \newline In terms of detecting biomarkers and life, the triple signature O$_3$, CO$_2$
and H$_2$O is a possibility to avoid false-positive detections of
biospheres, as proposed by, e.g., \citet{selsis2002}.
\citet{sagan1993} proposed O$_2$ (or its tracer O$_3$) and CH$_4$ as
combined biomarkers.


\section{Models and atmospheric scenarios}

\label{theatmosphere}

\subsection{Atmosphere model}

The spectra shown here have been calculated using atmospheric
profiles from scenarios summarized in \citet{vparis2010gliese}.
Atmospheric profiles (temperature, pressure, water) were calculated
using a 1D radiative-convective model. The model is originally based
on the model of \citet{kasting1984}. It solves the radiative
transfer equation to calculate temperature profiles in the
stratosphere. The stellar flux is treated in 38 spectral
intervals of varying width ranging from 0.237 to 4.545 $\mu$m. The radiative transfer
scheme uses a $\delta$-2-stream method \citep{Toon1989} to
incorporate Rayleigh scattering by N$_2$, CO$_2$ and H$_2$O
(\citealp{vardavas1984}, \citealp{vparis2010gliese}). Additionally,
absorption by H$_2$O and CO$_2$ is taken into account in the visible
and near-IR \citep{pavlov2000}. In the troposphere, the atmosphere
is assumed to be convective. Hence, temperature profiles are assumed
to be adiabatic, based on \citet{kasting1988} and
\citet{kasting1991}. The water profile is calculated using a fixed
relative humidity profile \citep{manabewetherald1967}. More details
on the model are given in \citet{vparis2008} and
\citet{vparis2010gliese}.

\subsection{Atmospheric scenarios}
\label{scenarios}

The model scenarios used the orbital distance of 0.22 AU and
eccentricity of $e$=0.38 for GL\,581\,d \citep{mayor2009gliese}. The
stellar input spectrum was based on a synthetic  NextGen spectrum
\citep{hauschildt1999} in the visible and near-IR merged with
measurements by the International Ultraviolet Explorer satellite in
the UV, as described by \citet{vparis2010gliese}.

Model CO$_2$ concentrations were chosen to be consistent with CO$_2$
concentrations on present Venus and Mars (95\%), present Earth
(3.55$\cdot$10$^{-4}$) as well as assumed scenarios of the early
Earth (5\%, e.g., \citealp{kasting1987}). Surface pressures were
chosen such that model column densities were comparable to scenarios
of early Earth or early Mars in the literature (e.g.,
\citealp{goldblatt2009faintyoungsun}, \citealp{tian2010}). Three
different atmospheric  scenarios were considered, defined by the
CO$_2$ concentration: low (355 ppm volume mixing ratio, vmr), medium
(5\% vmr) and high (95\% vmr) CO$_2$ cases. The model atmospheres
further contained water and molecular nitrogen as a filling gas,
i.e. the model atmospheres are CO$_2$-H$_2$O-N$_2$ mixtures. In each
of these cases, the surface pressure was varied from 1 to 20 bar. For the 1 and 20 bar cases, respectively, spectra are presented below. The surface albedo was kept constant at $A_{\rm{surf}}$=0.13 for all runs which is the measured surface albedo of Earth \citep{rossow1999}. In doing so, clouds were explicitly excluded in the climate calculations of \citet{vparis2010gliese}. Table \ref{listofruns} summarizes the atmospheric scenarios considered in this work.

\begin{table}[H]
  \caption{Atmospheric scenarios for GL\,581\,d considered here
  }\label{listofruns}
\begin{center}
\begin{tabular}{lcc}
 \hline
   Scenario  &   $p$ [bar]     & CO$_2$ vmr           \\
  \hline
    low-CO$_2$  & 1, 20       &3.55 $\cdot$ 10$^{-4}$           \\
    medium-CO$_2$ & 1, 20       &0.05                              \\
    high-CO$_2$& 1, 20       &0.95                       \\
\end{tabular}
\end{center}

\end{table}

The model calculations of \citet{vparis2010gliese} suggested that
four scenarios with high CO$_2$ partial pressures (high-CO$_2$ 5, 10
and 20 bar as well as medium-CO$_2$ 20 bar) were habitable, with
surface temperatures above 273 K, i.e. above the freezing point of
water. These results were in broad agreement with results from
\citet{wordsworth2010}, \citet{hu2011} and \citet{kaltenegger2011}.

\subsection{Computation of spectra and signal-to-noise ratios}

\label{specmodel}

\begin{figure*}
 \resizebox{\hsize}{!}{ \includegraphics[width=220pt]{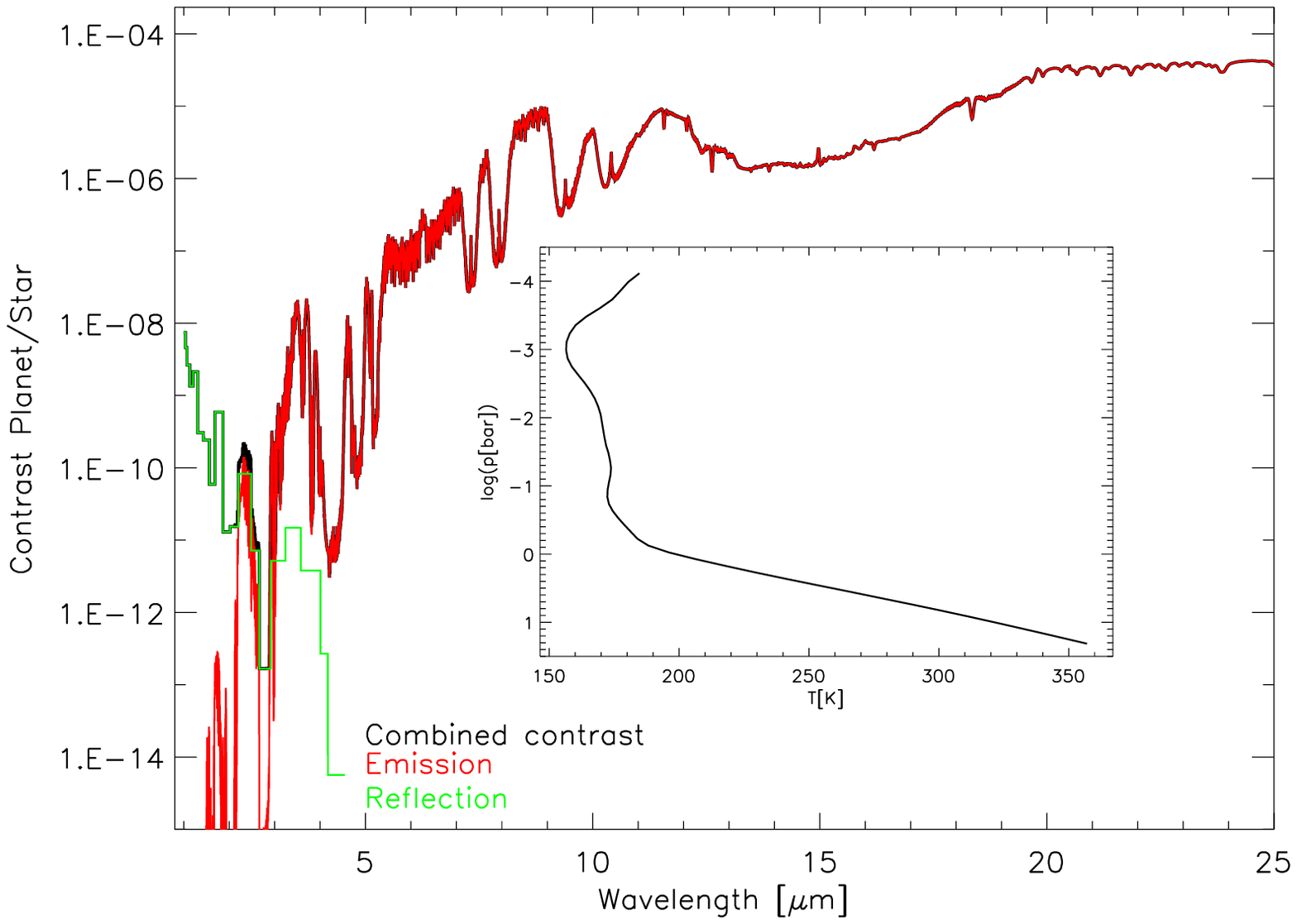}
 \includegraphics[width=220pt]{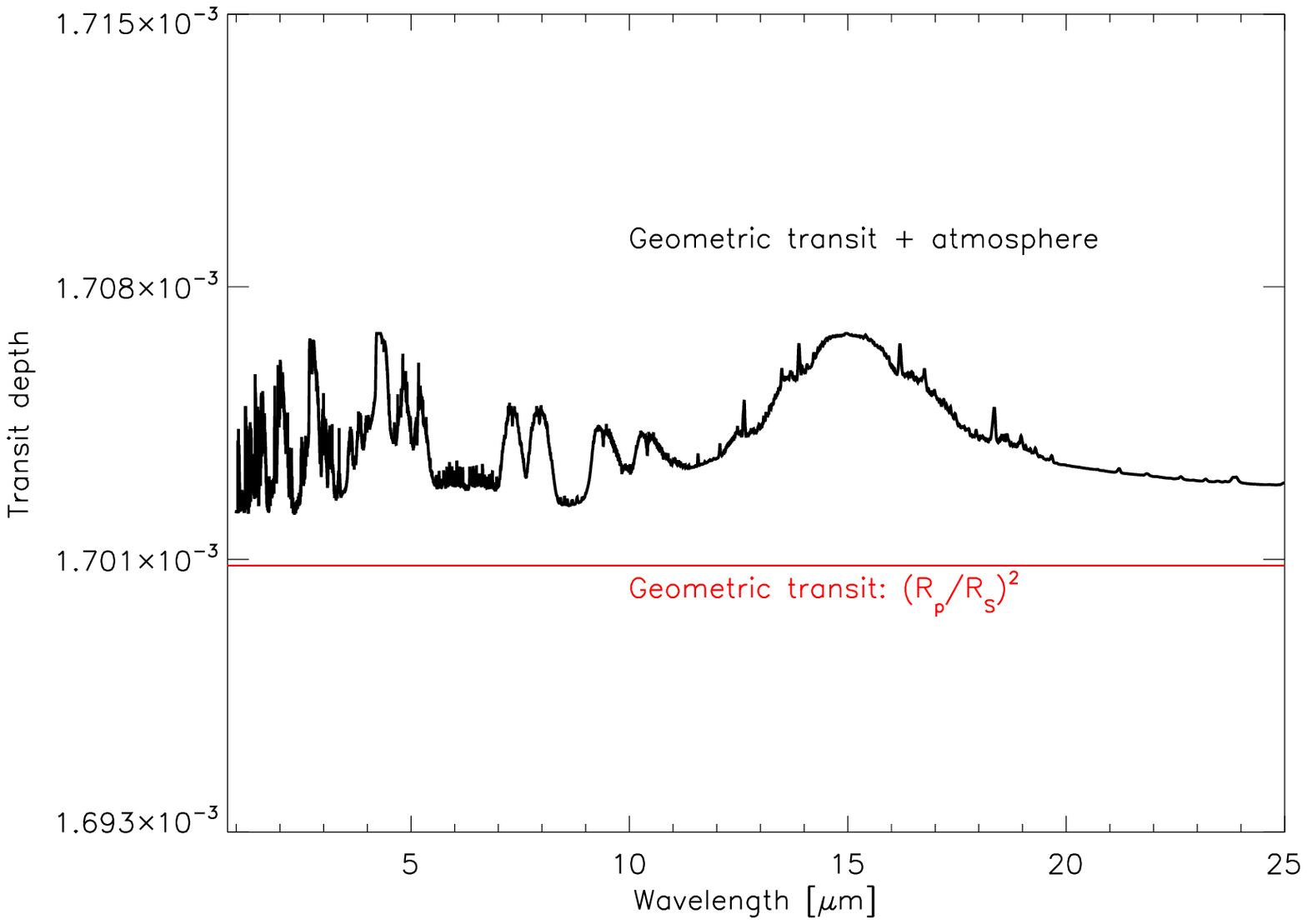} }
  \\
  \caption{Spectra of the 20 bar high-CO$_2$ case. Left: Secondary eclipse spectrum. Inlet shows
   the corresponding temperature-pressure profile.  Right: Transit depth spectrum.}
  \label{example_emission}
\end{figure*}

The line-by-line code MIRART-SQuIRRL \citep{schreier2003} was used
for the calculation of high-resolution synthetic planetary emission,
brightness temperature and transmission spectra. Note that the collision-induced absorption (CIA) of CO$_2$ is not included here, although the CIA is thought to be important in some spectral regions around 7\,$\mu$m and longwards of 20\,$\mu$m (e.g.,  \citealp{wordsworth2010cont}). \newline Furthermore, the output of the stellar radiative transfer code of the
climate model (see above) was used to produce reflection spectra
$I_R$:

\begin{equation}\label{reflexion}
    I_R=\frac{A_S}{2}\cdot I_S\cdot \frac{R_s^2}{d^2}
\end{equation}

where $I_S$ is the stellar spectrum, $A_S$ the spectral
albedo from the stellar radiative transfer code, $R_s$ the stellar
radius and $d$ the orbital distance. This approach of using the
climate model output to construct spectra is based on
\citet{kitzmann2011}. A similar method was used by
\citet{wordsworth2011} to calculate synthetic broadband emission spectra of GL
581 d model atmospheres.

These spectra were then used to calculate contrast spectra
(i.e., emission + reflection) as well as spectra of
effective tangent height of the atmosphere. Calculations in our work closely follow
\citet{rauer2011} where more details can be found. The spectra were
calculated on an equidistant spectral grid (in wavenumber), hence
the spectral resolution $R$ varies between
$R\approx$2-10$\cdot$10$^3$, depending on wavelength.

For a detection of a spectroscopic feature, the relevant quantity is
the signal-to-noise ratio (SNR). In contrast to the SNR
calculations by, e.g., \citet{Kaltenegger2009} or \citet{rauer2011},
we take into account not only the stellar photon noise, but also the thermal emission
of the telescope, the zodiacal emission and the dark noise. The spectrum of GL 581 used in the SNR calculations is the synthetic spectrum described above. We base our telescope parameters on the James Webb Space
Telescope (JWST), assuming a 6.5 m aperture and a detection
efficiency of 0.15 \citep{Kaltenegger2009}. More details on
the noise estimates can be found in the Appendix \ref{appendix_noise}. The fictitious
transit duration of GL\,581\,d, hence the assumed integration time, is
calculated to be 4.15 hours. SNRs are calculated for a spectral
resolution of $R$=10, which is a reasonable value in the context of
exoplanet characterization.

\begin{figure*}
 \resizebox{\hsize}{!}{ \includegraphics[width=220pt]{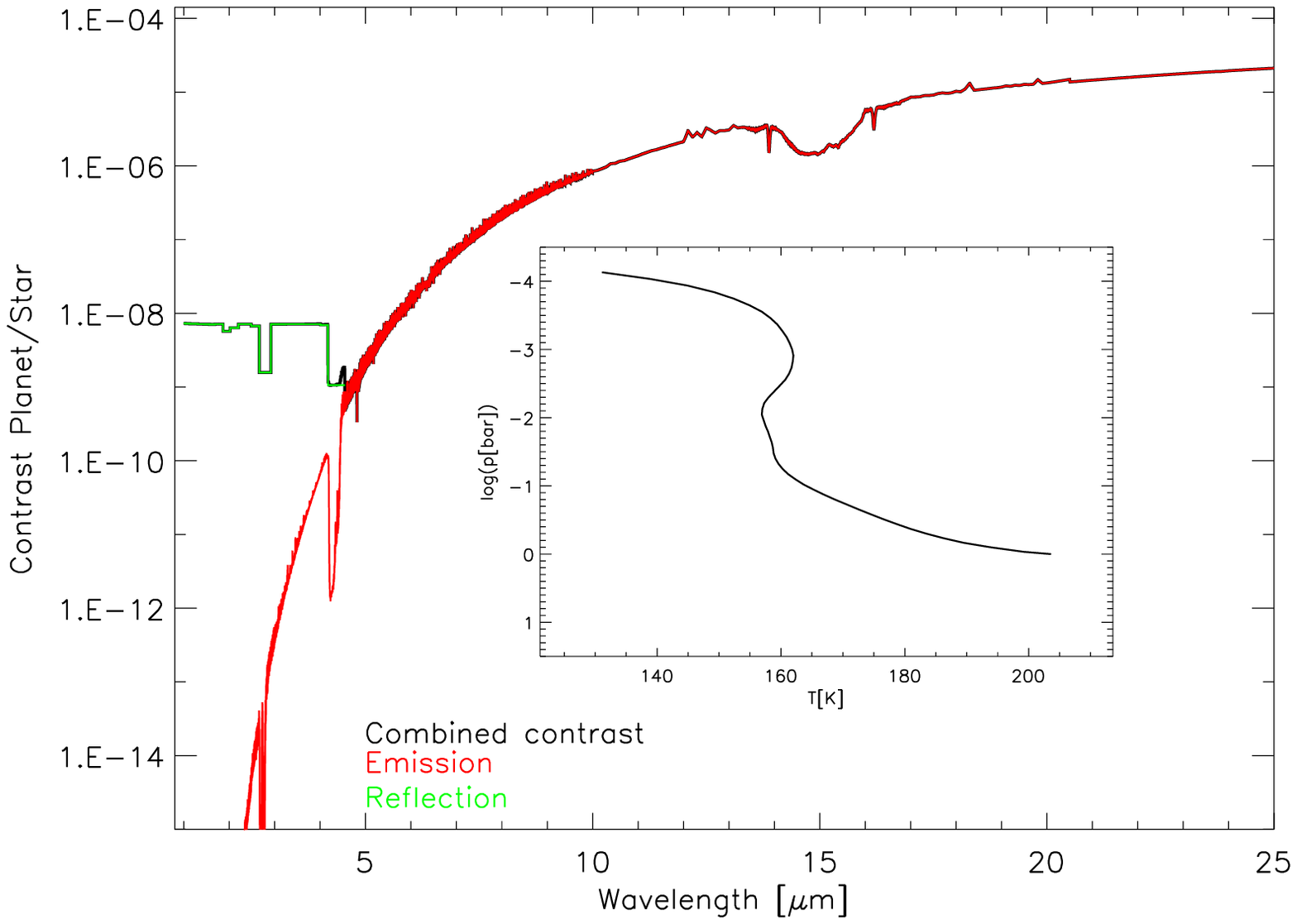}
 \includegraphics[width=220pt]{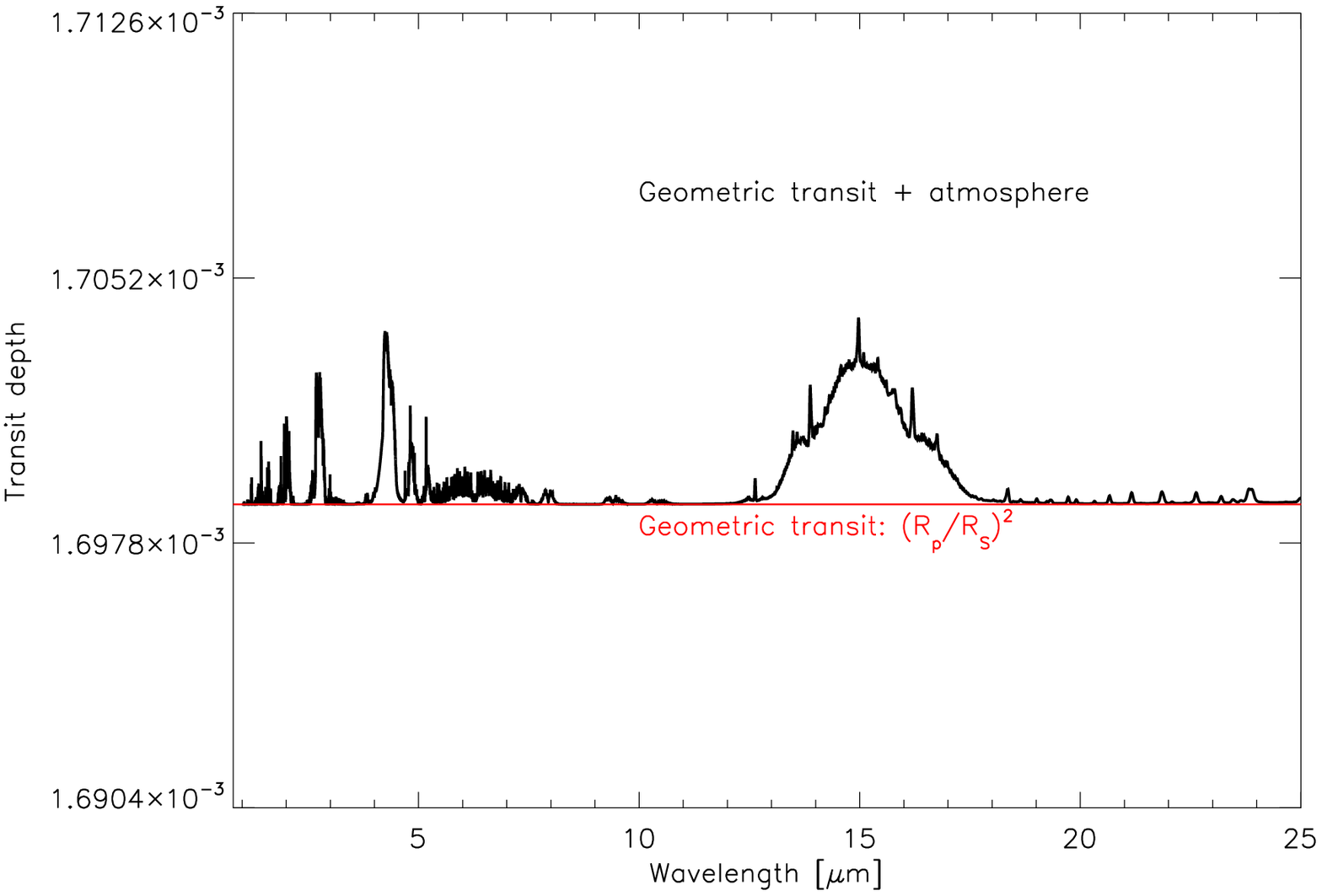}}
  \\
  \caption{Same as Fig. \ref{example_emission}, but for the 1 bar low-CO$_2$ case.}
  \label{example_low_1bar}
\end{figure*}

\begin{figure*}
 \resizebox{\hsize}{!}{ \includegraphics[width=220pt]{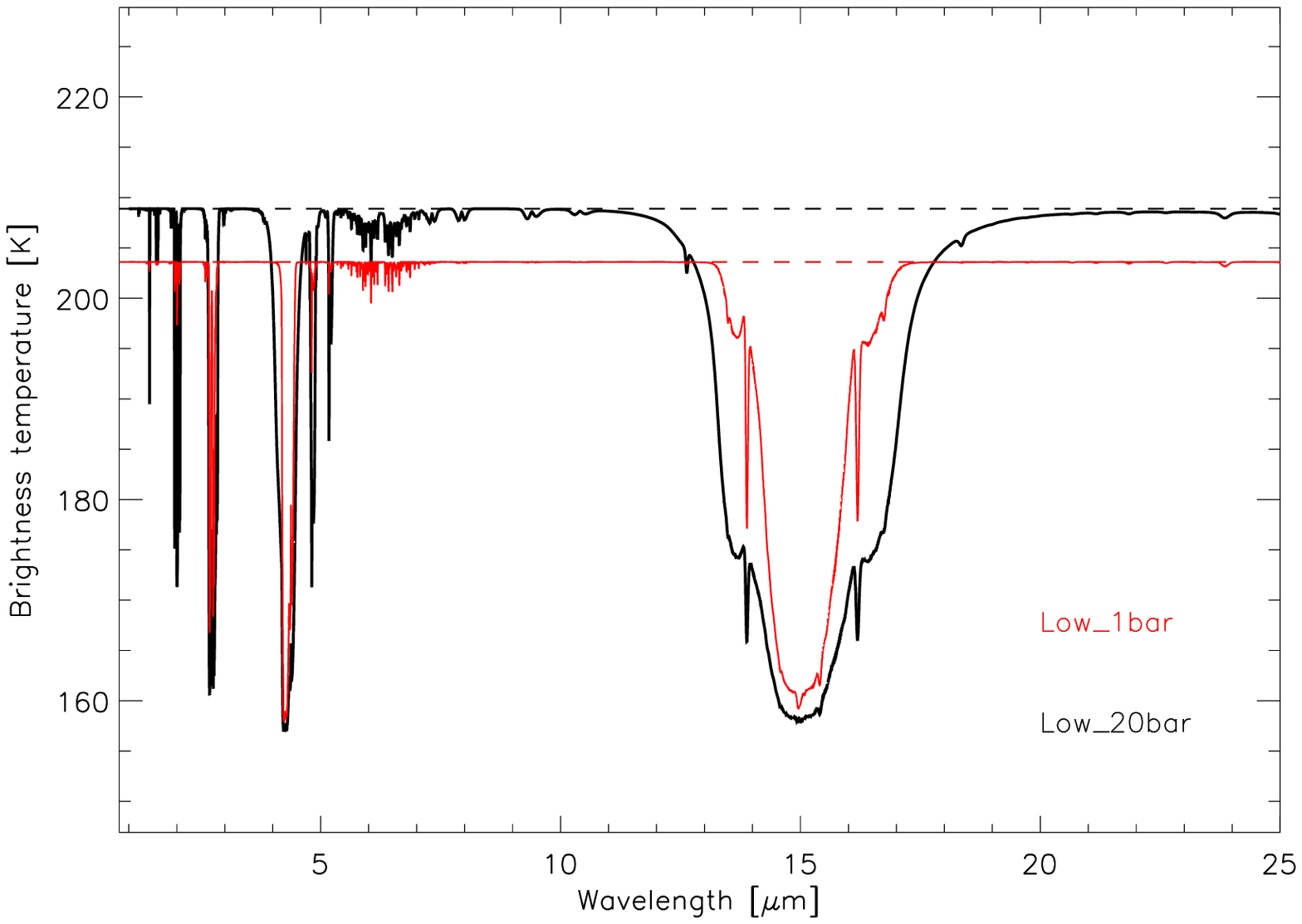}
  \includegraphics[width=220pt]{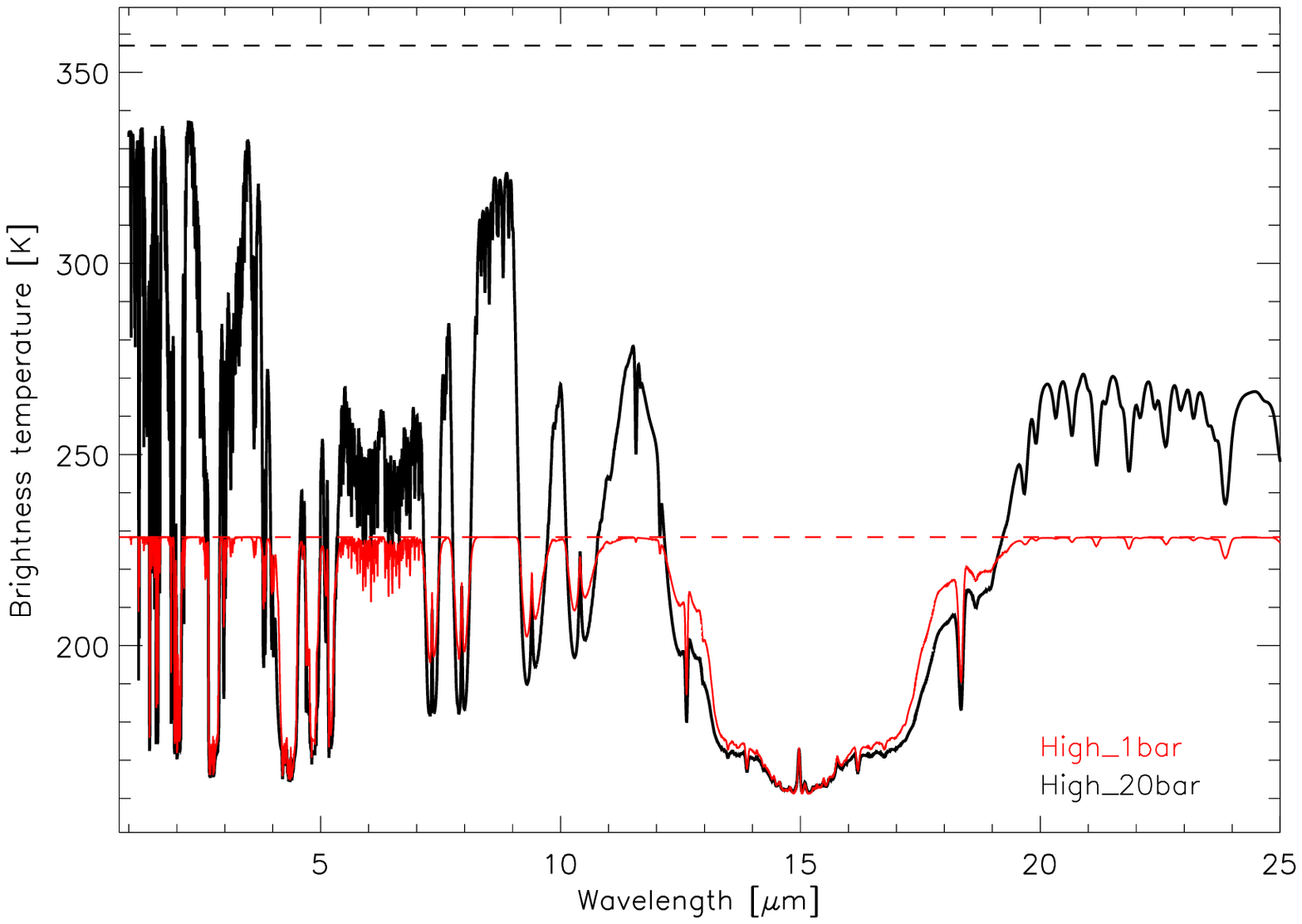}}
 \\
  \caption{Brightness temperature spectra (based on planetary emission only) of the 1 and 20 bar scenarios.  Left: Low-CO$_2$ case. Right: High-CO$_2$ case. Surface temperatures are indicated by horizontal dashed lines.} 
 \label{example_bright}
\end{figure*}

\section{Results and discussion}

\label{resultsect}

\subsection{Presence of an atmosphere}

The left part of Fig.
\ref{example_emission} shows the contrast spectrum of the
high-CO$_2$ 20 bar case. It is clearly seen that the reflection
component dominates the near-IR up to about 2.5 $\mu$m and in the window between 3-4 $\mu$m . For longer
wavelengths, the emission of the planet is the main
component. The broad water and CO$_2$ absorption bands are clearly seen in the spectrum. Hence, the
existence of an atmosphere as well as the presence
of water and CO$_2$ could be inferred. Interestingly, the
planet-star contrast is rather low, even though GL 581 is an M-type
star and GL\,581\,d a Super-Earth. The contrast reaches about
4$\cdot$10$^{-5}$ in the mid-IR which is about an order of magnitude
higher than the contrast between Earth and the Sun. However, it is
about 100 times lower than corresponding values for hot Jupiters. The right part of Fig. \ref{example_emission} shows the synthetic transit depth spectrum for the 20 bar high-CO$_2$
case. It can be seen that due to the presence of large amounts of
water and CO$_2$, the planet appears larger than its geometric
radius at all IR wavelengths. Hence, the presence of an atmosphere could also be clearly inferred from transmission spectra due to the wavelength-dependent apparent radius.

Figure \ref{example_low_1bar} also shows contrast and transit spectra, but for the 1 bar low-CO$_2$ case. The spectra are relatively flat, except for the strong CO$_2$ fundamental bands. In these bands, the presence of an atmosphere could be inferred, as for the 20 bar high-CO$_2$ case. Since the atmosphere is very dry (partial pressure of water less than 10$^{-5}$ bar), the water bands are difficult to discern in the emission spectrum. In contrast to the high-CO$_2$ 20 bar case, the secondary eclipse spectrum of the low-CO$_2$ 1 bar case is dominated by the reflection component up to 4.5 $\mu$m, i.e. the limit of the stellar radiative transfer code of the atmospheric model.

\subsection{Atmospheric characterization}

\begin{figure*}
  \resizebox{\hsize}{!}{\includegraphics[width=320pt]{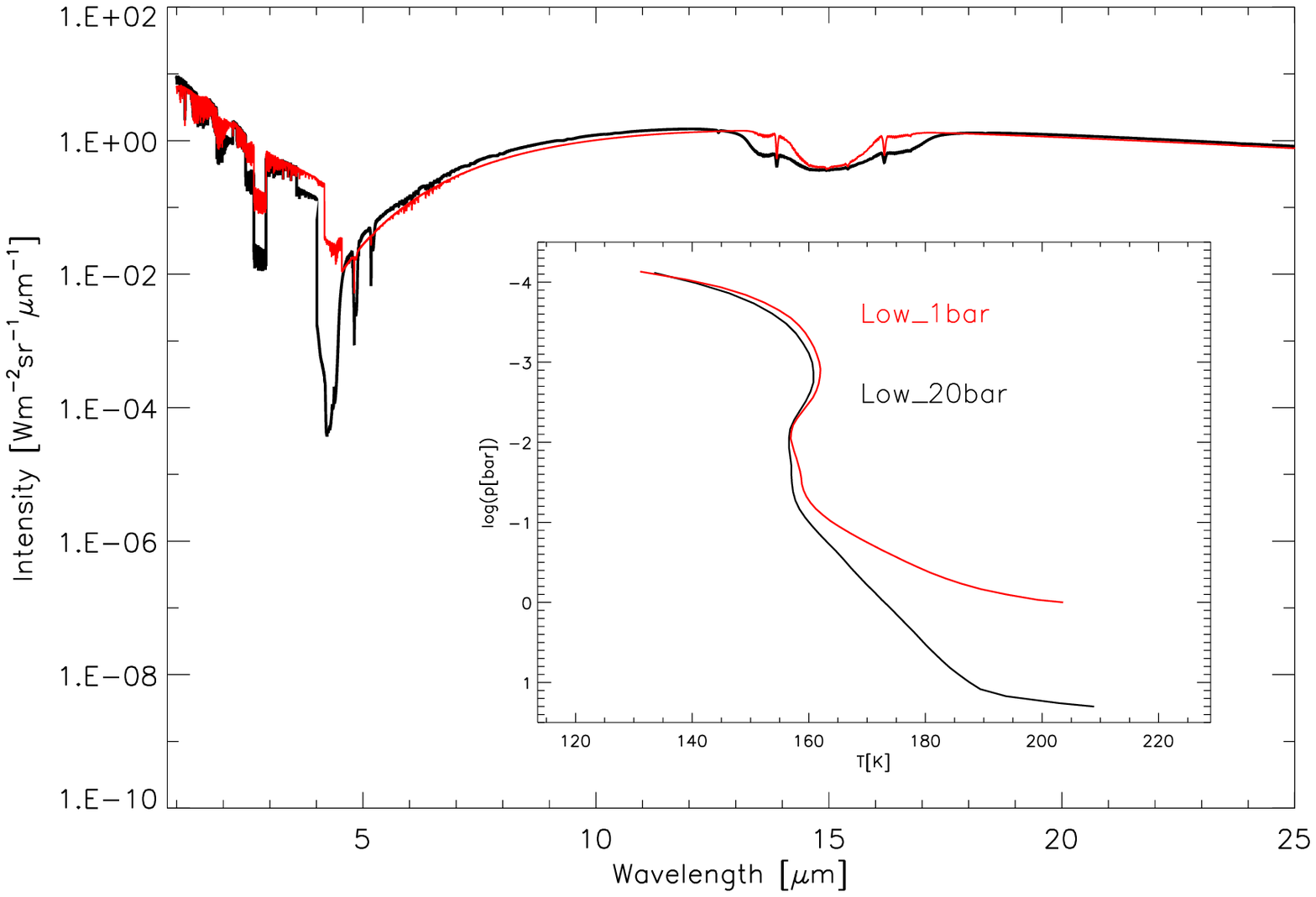}
  \includegraphics[width=320pt]{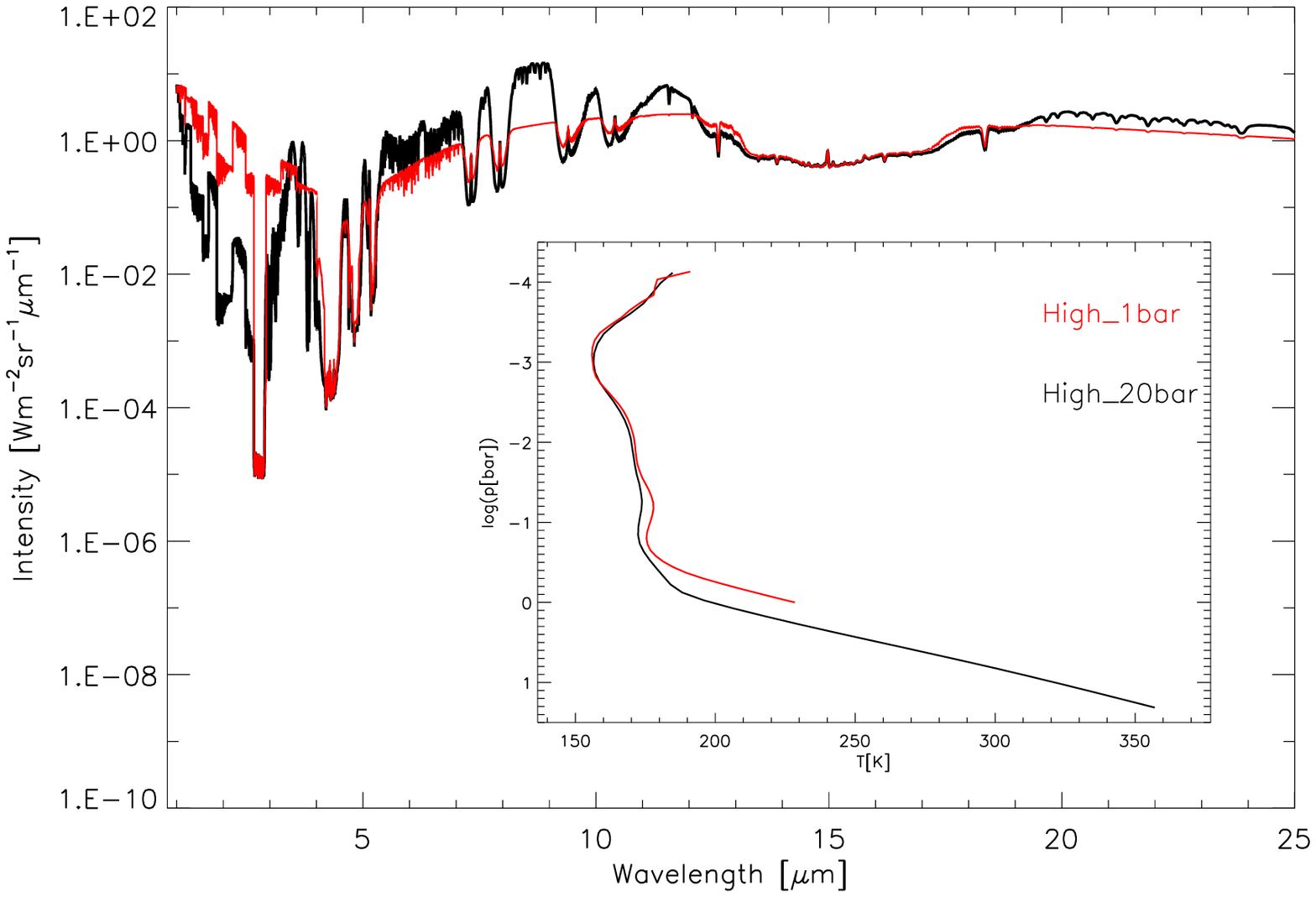}}
  \\
  \caption[Comparison of planetary spectra of different
  scenarios: Pressure effect]{Comparison of secondary eclipse spectra of different
  scenarios: Pressure effect. Scenarios as indicated. Inlet shows
   the corresponding temperature-pressure profiles. }
   \label{pressure_effect}
\end{figure*}

\begin{figure*}
 \resizebox{\hsize}{!}{ \includegraphics[width=220pt]{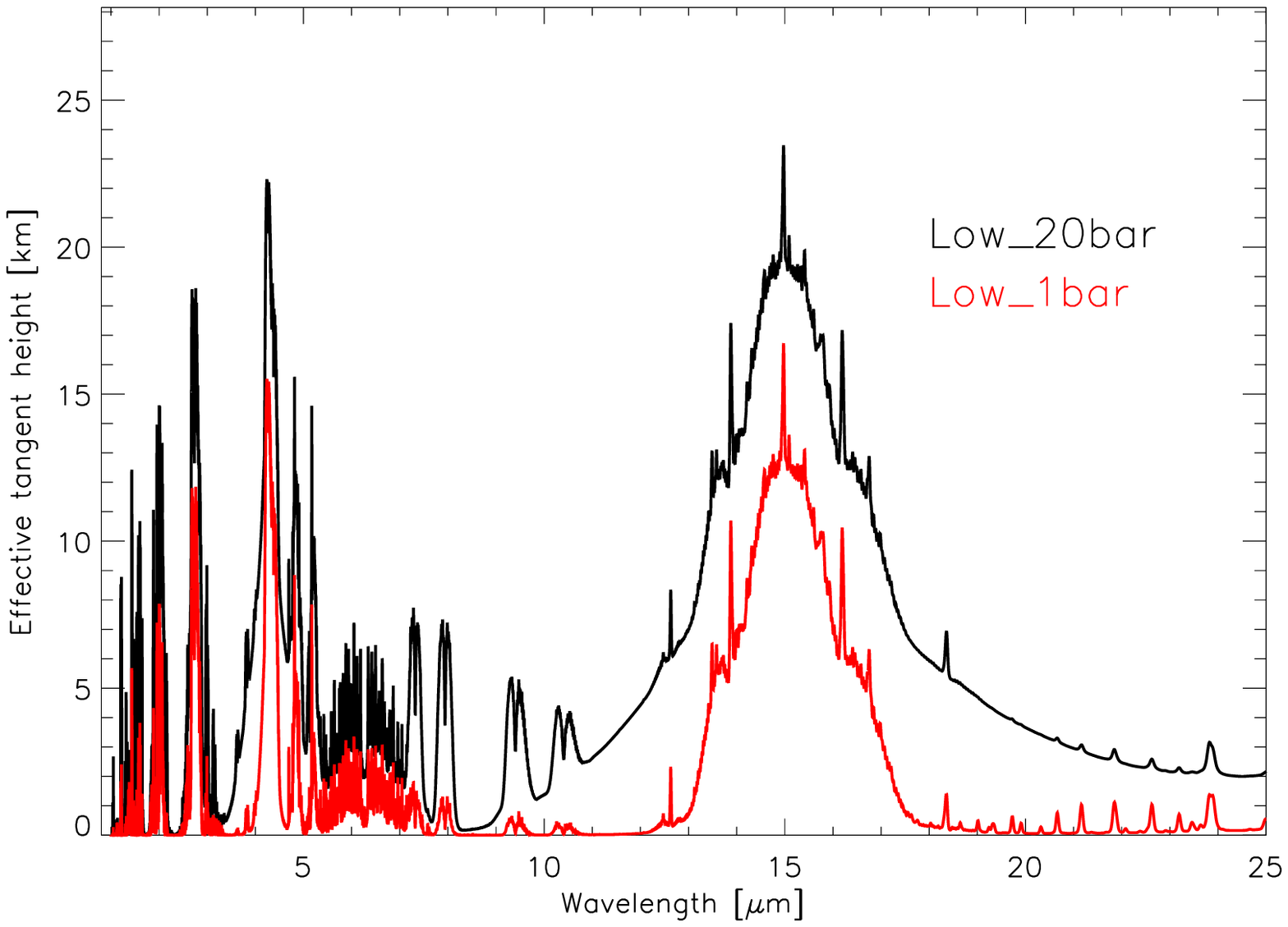}
     \includegraphics[width=220pt]{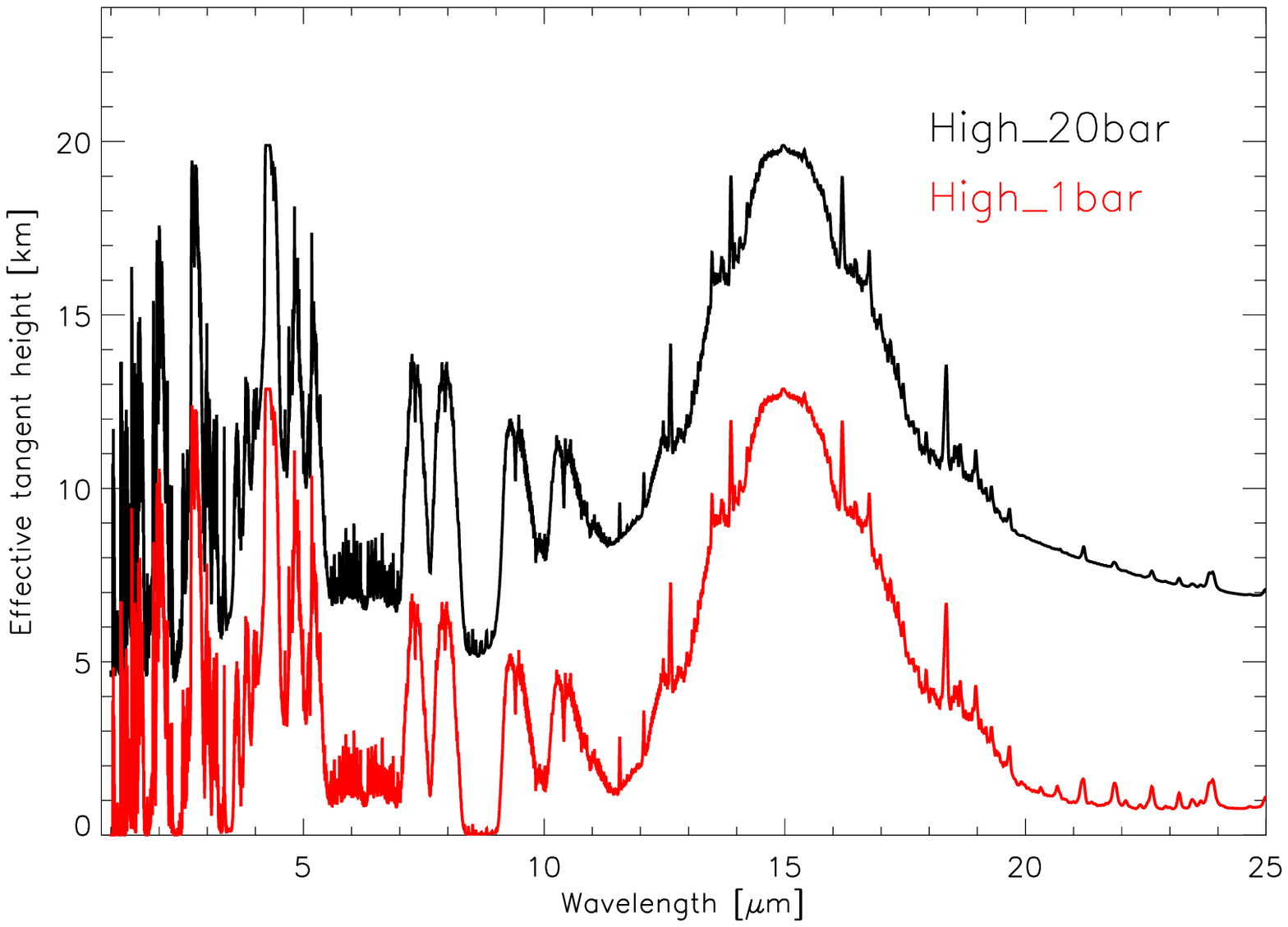}}
 \\
  \caption[Transmission spectra: Pressure effect]
  {Transmission spectra: Pressure effect.
  Low-CO$_2$ (left) and high-CO$_2$ (right).}
 \label{trans_pressure}
\end{figure*}

After securely detecting the atmosphere, the next step would
be its characterization (i.e., composition, surface pressure) and,
from there, assessing the surface conditions, hence potential
habitability.

\subsubsection{Surface temperature}

Figure \ref{example_bright} shows the brightness temperature spectra of the 1 and 20 bar scenarios of the low-CO$_2$ (left) and the high-CO$_2$ case (right).  Note that brightness temperature spectra are based only on the emission spectrum since brightness temperatures calculated from the contrast due to the reflection spectrum would yield values up to
about 700 K (near 1 $\mu$m). Therefore, characterization of atmospheric temperatures  of terrestrial planets 
is not possible in the near-IR up to about 4-5 $\mu$m. \newline The left part of Fig. \ref{example_bright} shows that for the low-CO$_2$ scenarios, the surface temperature could be inferred from the brightness temperature spectra since the atmosphere is transparent except in the CO$_2$ fundamental bands. By contrast, 
as can be seen in the right part of Fig. \ref{example_bright}, the
difference between the brightness temperature and the surface
temperature is always non-zero in the high-CO$_2$ 20 bar case. This
means that the emission spectrum does not allow for a determination
of the surface temperature, hence to assess potential habitability.
The reason for this is that the atmosphere is optically thick for
thermal radiation due to the large amounts of CO$_2$ and water in
the atmosphere \citep{vparis2010gliese}. In the 1 bar high-CO$_2$ case, some spectral windows would still allow for the determination of the surface temperature which is mostly due to the fact that the atmosphere is much drier than in the high-CO$_2$ 20 bar case.

\subsubsection{Surface pressure}

In Fig. \ref{pressure_effect}, the 1 and 20 bar runs with high and
low CO$_2$ concentrations are compared to each other to illustrate
how surface pressures could be inferred from secondary
eclipse spectra. In the low-CO$_2$ case (left), the main effect can
be seen in the 15 $\mu$m band, which is considerably broader for the
20 bar run than for the 1 bar run. This is simply due to the fact
that the line center becomes optically thick at pressures of about
100 mbar, whereas the line wings are transparent up to pressures of
the order of 5-10 bar. Also, the 2.7 and 4.3\,$\mu$m bands are much deeper in the 20 bar scenario compared to the 1 bar scenario. In the high-CO$_2$ case (right), the pressure
effect in the 15 $\mu$m CO$_2$ band is much less pronounced. Line
wings are already saturated at pressures well below 1 bar. Due to
the large difference in surface temperature (see inlet), some
spectral regions (e.g., windows near 8-9 $\mu$m or around 11 $\mu$m) differ considerably. Furthermore, in the near-IR, the reflection spectra show large differences, owing to the strong decrease of the spectral albedo with surface pressure.  This demonstrates the use of near-IR secondary eclipse measurements for determining atmospheric characteristics besides temperature.

Figure \ref{trans_pressure} shows the effect of changing surface
pressure on the transmission spectrum for the low and the
high-CO$_2$ 1 and 20 bar cases. The spectra show significant
differences. For example, in the 9.5 $\mu$m CO$_2$ band, tangent
heights differ by about a factor of 2 to 3. However, in terms of
absolute height, this amounts to 5-10 km at most.
Note that the CO$_2$ bands at 7-10 $\mu$m are visible already in the
spectra of the high-pressure low-CO$_2$ scenarios, contrary to the
emission spectra where these bands did not appear (see Fig.
\ref{pressure_effect}).

These results imply that transmission spectra are in general more
sensitive to surface pressure than emission spectra for the cases
studied. This is due to two main reasons. Firstly, the tangent
height to first order depends on the atmospheric scale height,
$H\sim$ $T/m_a$ ($T$ is the surface temperature and $m_a$ the mean
molecular weight of the atmosphere). Thus, for higher surface
pressures, and corresponding higher surface temperatures, scale
heights are larger. Secondly, for higher
surface pressures, atmospheres extend further out to space. For
example, the 20 bar high-CO$_2$ case has its model lid
(corresponding to a pressure of 6.6$\cdot$10$^{-5}$ bar) at 20 km,
whereas the 1 bar case has its model lid at 13 km altitude, which
corresponds roughly to 3 scale heights difference.

\subsubsection{Surface albedo}

The reflectivity of the surface, the surface albedo, is an important parameter to distinguish between different types of surfaces, such as oceans or ice. Reflection spectra could in principle reveal information about the value of the surface albedo. \newline However, for optically thick atmospheres, as already discussed above, the reflection 
spectrum does not yield information on the surface albedo in the high-CO$_2$ 20 bar case. Without the influence of the atmosphere, the reflection contrast would be 7.1 $\cdot$10$^{-9}$ (with a fixed surface albedo of  $A_{\rm{surf}}$=0.13, see Sect. \ref{scenarios}).  As can be seen from Fig. \ref{example_emission}, the reflection contrast is far lower than this value, indicating that even cloud-free atmospheres could inhibit surface characterization via reflection spectra. The same is true for the high-CO$_2$ 1 bar and the low-CO$_2$ 20 bar cases (not shown). Only in the low-CO$_2$ 1 bar case, the reflection spectrum would allow to infer the surface albedo (see Fig. \ref{example_low_1bar}).

\begin{figure}[H]
 \resizebox{\hsize}{!}{ \includegraphics[width=220pt]{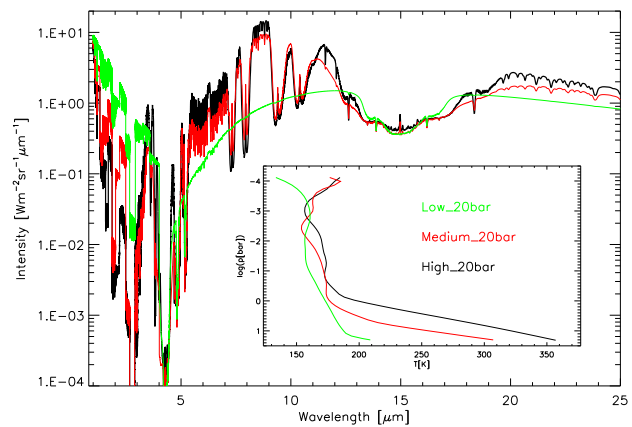}
}
  \\
  \caption[Comparison of planetary spectra of different
  scenarios: Concentration effect]{Comparison of secondary eclipse spectra of different
  scenarios: Concentration effect. Inlet shows
   the used temperature-pressure profiles.}\label{concentration_effect}
\end{figure}

\subsubsection{Atmospheric composition}

Fig. \ref{concentration_effect} shows secondary eclipse  spectra at equal surface
pressures (20 bar), but for different CO$_2$ concentrations (high,
medium, low). It is rather difficult to distinguish between the
medium and high-CO$_2$ scenario, except for some narrow spectral windows. When comparing Fig.
\ref{pressure_effect} with Fig. \ref{concentration_effect}, it is
therefore difficult to decide whether the shape of the spectrum is
actually due to a pressure difference at high CO$_2$ concentration
or a concentration difference at high surface pressures. It is,
however, possible to distinguish between high or low CO$_2$
concentrations by the presence of many weak CO$_2$ bands in the
high-CO$_2$ case, e.g. at around 7, 9 and 10 $\mu$m. These bands do
not appear in the spectrum unless CO$_2$ concentrations exceed
several percent. Additionally, the presence of water can be inferred from the strong 6.3 $\mu$m fundamental band and the signatures of the rotation band in the mid- to far-IR.  Thus, dry and moist atmospheres could be distinguished.

\begin{figure}[H]
 \resizebox{\hsize}{!}{ \includegraphics[width=220pt]{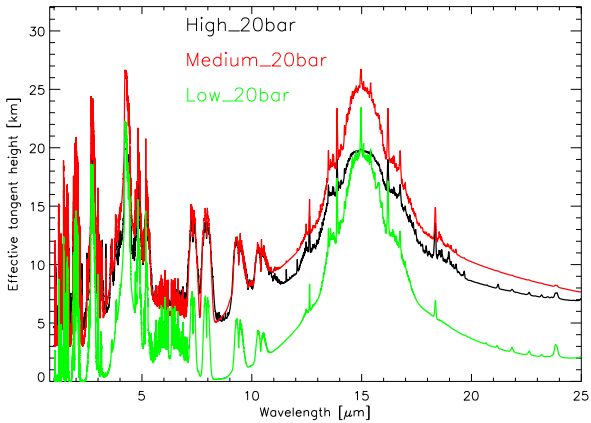}
}
 \\
  \caption[Transmission spectra: Concentration effect]
  {Transmission spectra: Concentration effect.}
 \label{trans_conc}
\end{figure}

In order to show the effect of changing CO$_2$ concentration on
transmission spectra, Fig. \ref{trans_conc} compares the 20 bar
scenarios for low, medium and high CO$_2$ concentrations.
Interestingly, the medium-CO$_2$ case shows a more pronounced 15
$\mu$m band of CO$_2$ despite the fact that the high-CO$_2$ scenario
is significantly warmer at the surface (by about $\sim$ 40 K). This is due to the
lower mean atmospheric molecular mass (hence, higher scale height)
which is $\sim$ 29 g mol$^{-1}$ in the medium-CO$_2$ case and $\sim$
43 g mol$^{-1}$ in the high-CO$_2$ case. The weak bands around 7 and
10 $\mu$m do not differ by much. Still, in the water rotation bands
(longwards of 20 $\mu$m) and the strong CO$_2$ fundamentals, rather
large differences in the spectrum can be seen. This indicates that
transmission spectroscopy could be able to distinguish between
different CO$_2$ concentrations.

\begin{figure*}
  \resizebox{\hsize}{!}{\includegraphics{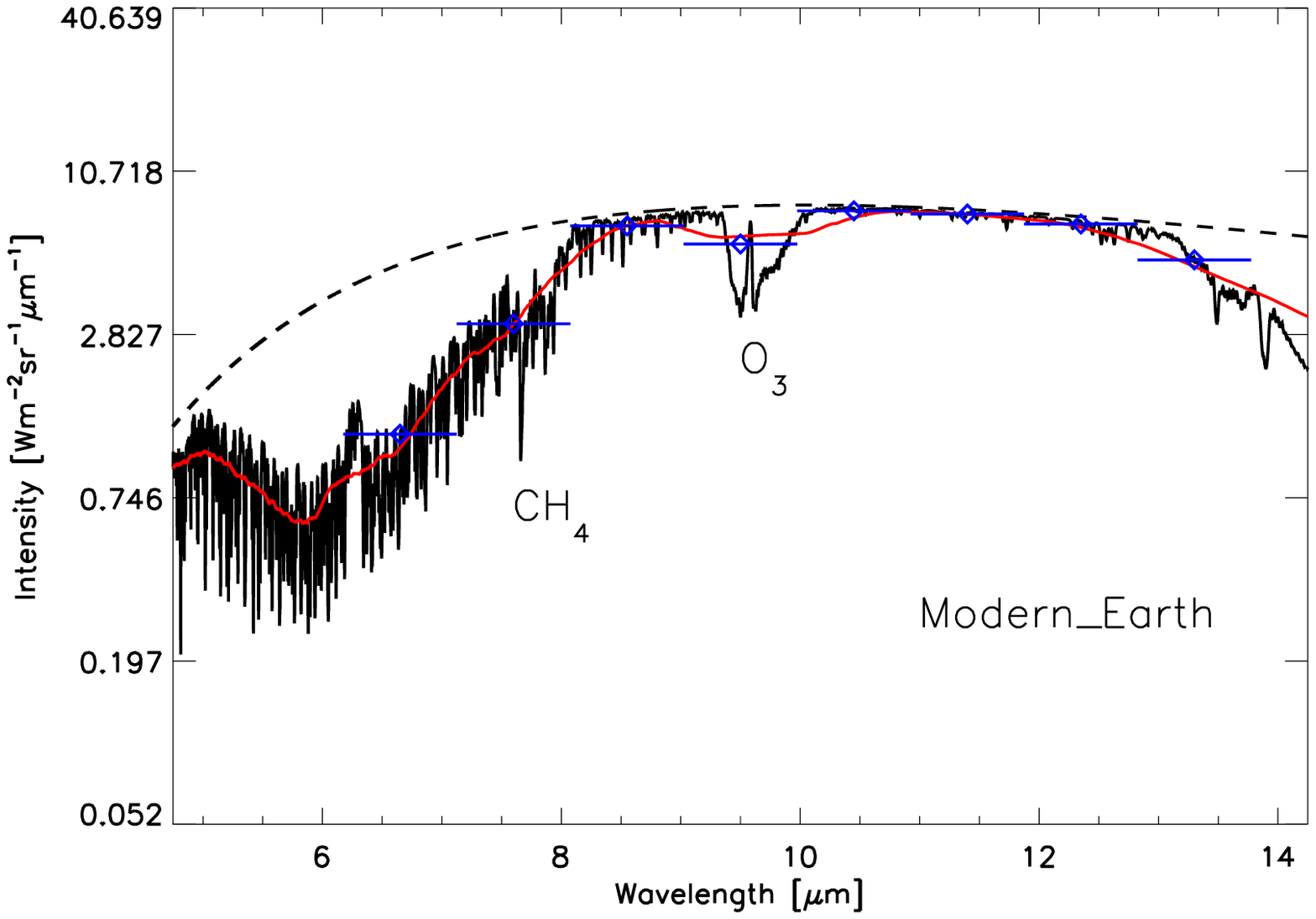}
  \includegraphics{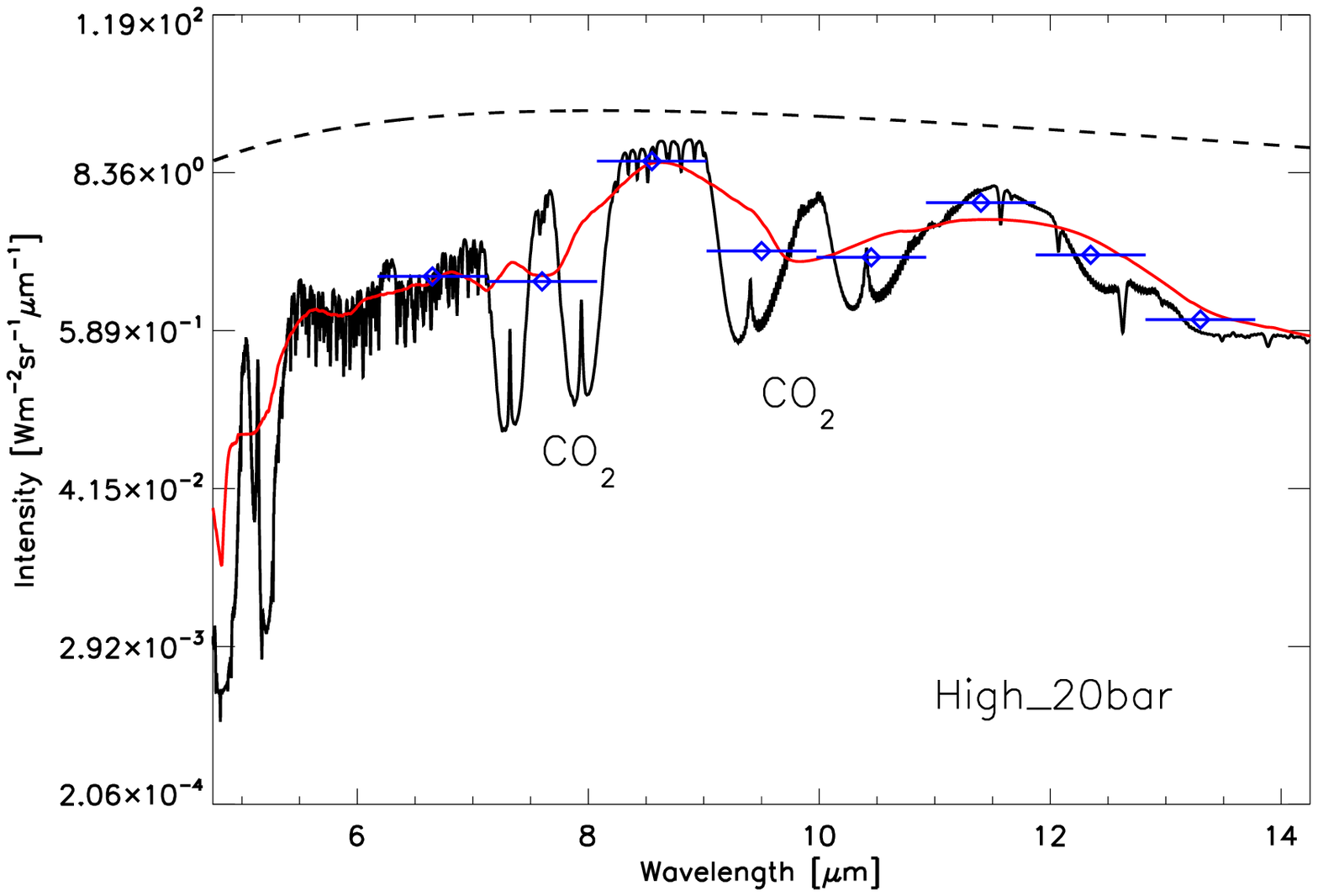}
  }\\    
  \resizebox{\hsize}{!}{\includegraphics{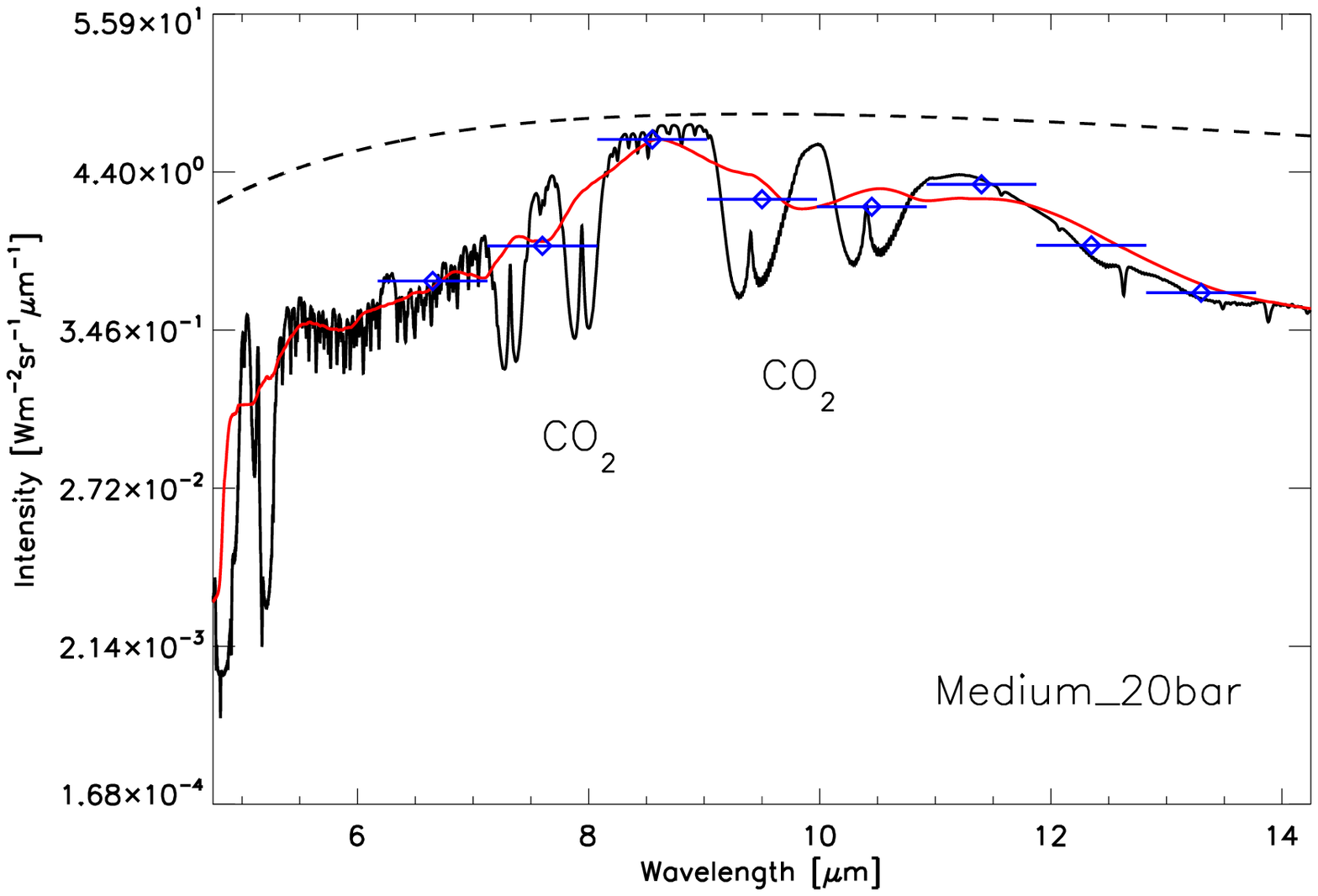}
  \includegraphics{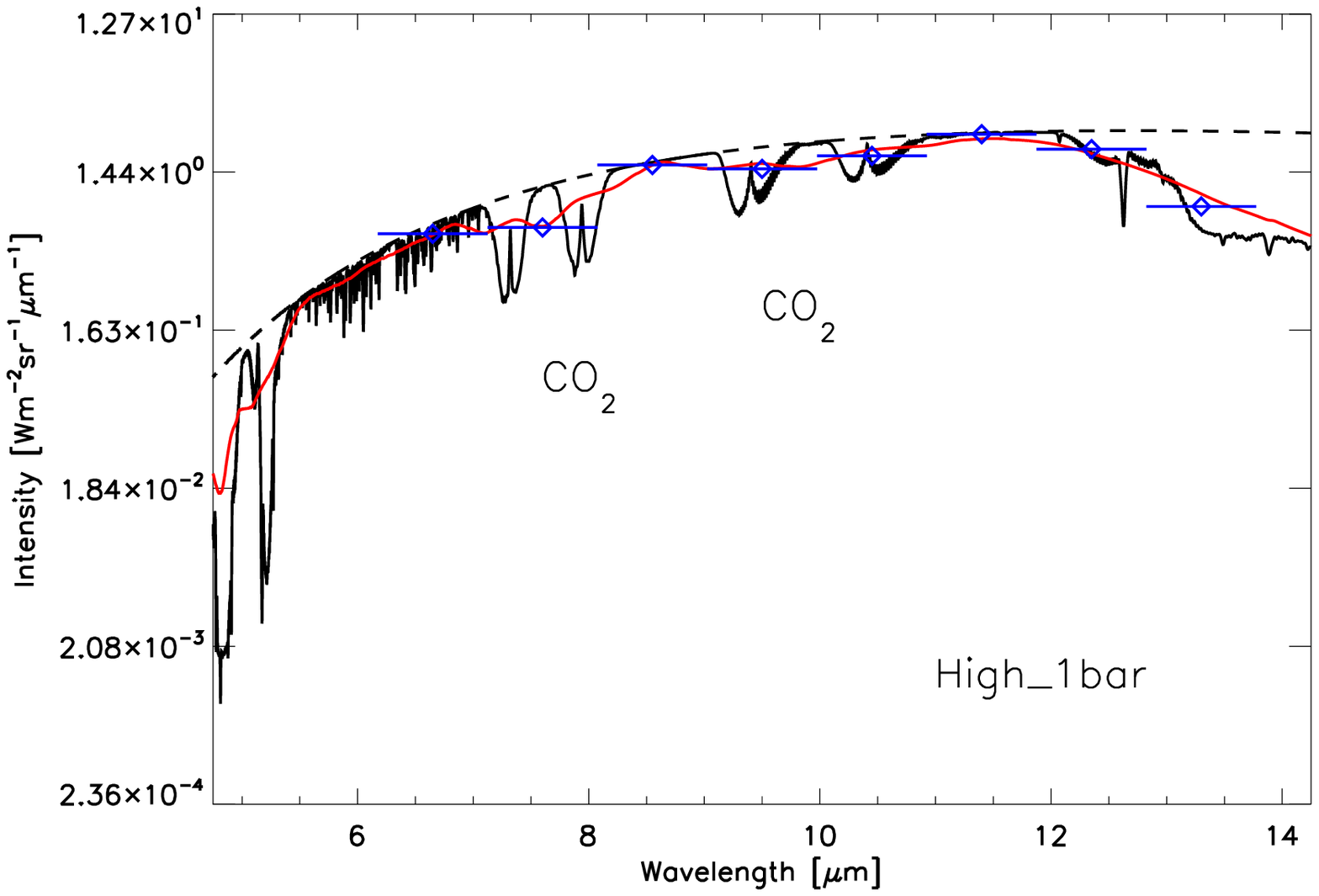}
  }
 \\
  \caption[Possibility of a false-positive biomarker detection]
  {Possibility of a false-positive biomarker detection for different
  scenarios, as indicated. Modern Earth spectrum taken from \citet{grenfell2011}. Planck curves of corresponding surface temperatures as dashed lines.
 Smoothed (red) and binned spectra (blue) are for a spectral resolution of 10.}\label{false_positive}
\end{figure*}

The same is true for the comparison of low and high-CO$_2$
scenarios. Here, however, the differences in the spectrum are
visible in the weak bands of CO$_2$ (e.g., around 10 $\mu$m) rather
than in the strong fundamentals (4.3 and 15 $\mu$m).

\subsubsection{Surface conditions}

In summary, the characterization of surface conditions on GL\,581\,d
by secondary eclipse spectroscopy would be rather difficult, even if the
planet were to be transiting. It would be possible to approximately
constrain CO$_2$ and water concentrations in the atmosphere,
however, constraining surface pressures or surface temperatures,
hence assess habitability, is complicated by degeneracies, as stated
above. Overall, results imply that it is easier to characterize the
atmospheric scenarios of GL\,581\,d with transmission spectroscopy
than with emission spectroscopy, especially with respect to the
habitable scenarios with massive CO$_2$ atmospheres. However,
surface pressures cannot be inferred directly with transmission spectra  since effective
tangent heights are always of the order of a few kilometres. Still, when combining several spectral bands and using both transmission and emission spectra, it may be possible to constrain the surface pressure as well as
CO$_2$ concentrations and the presence of water. Thus, in principle,
through atmospheric modeling surface conditions could be assessed.

\subsection{Biomarkers}

Although the model scenarios used in this work only consider
H$_2$O-CO$_2$-N$_2$ atmospheres, we can compare the computed spectra
with spectral signatures of modern Earth to discuss the potential
for false-positive or false-negative identifications of biomarkers.

\subsubsection{False-positive detections of biomarkers}

As an illustration, Fig. \ref{false_positive} shows high-resolution
and binned (with $R$=10) emission spectra of GL\,581\,d in comparison
to modern Earth. The modern Earth spectrum (taken from \citealp{grenfell2011})
is shown in the upper
left, the GL\,581\,d scenarios from this work are (clockwise) high-CO$_2$ 20 bar,
high-CO$_2$ 1 bar and the medium-CO$_2$ 20 bar case. The spectra are
centered around 9.6 $\mu$m which is the position of the main
absorption band of ozone. This absorption feature is close to an
absorption band of CO$_2$ at around 9.5 $\mu$m which is clearly seen
in all GL\,581\,d spectra in Fig. \ref{false_positive}.

Additionally indicated is the position of the 7.7 $\mu$m band of
methane (overlapping the 7.8 $\mu$m band of nitrous oxide) which
again is close to absorption bands of CO$_2$ at 7.5 and 8 $\mu$m.
The methane band is hardly discernible in the Earth spectrum, at the
considered resolution of $R$=10. However, in the GL\,581\,d scenarios
shown in Fig. \ref{false_positive}, an absorption is seen, which in
this case is due to CO$_2$. These CO$_2$ bands at biomarker
positions (7.5, 8, 9.5 $\mu$m) are also present in the transmission
spectra (e.g., Fig. \ref{trans_conc}), implying that transmission
spectra also suffer from the possibility of false-positive biomarker
detections.

At this low spectral resolution, the spectral features around 9.6
$\mu$m, due to CO$_2$ in the GL\,581\,d cases or due to ozone in the
Earth case, look similar. Thus,  the CO$_2$ bands could be mistaken for actual biomarker signals, at the
expected low spectral resolution used for exoplanet
characterization. This is the case for the emission spectra of
medium and high-CO$_2$ scenarios, whereas in the low-CO$_2$ case
(not shown in this Fig.), the respective CO$_2$ bands are too weak
to produce false detections. In the transmission spectra (e.g., Fig.
\ref{trans_conc}), these bands also appeared for high-pressure,
low-CO$_2$ scenarios which implies that false-positives are possible
in transmission spectra even for these scenarios.

One possible way of avoiding such false-positive detections in the
cases presented here (i.e., CO$_2$-rich atmospheres) is to exploit
the double nature of the CO$_2$ bands around 7 and 10 $\mu$m. If
spectral observations are performed, e.g., at 9.5 and 10.5 $\mu$m,
and both filters show a deep absorption, then the spectral
signatures are most likely due to CO$_2$. Hence, a possible spectral
characterization with respect to biomarkers should be done in all
main IR CO$_2$ bands (2-15 $\mu$m) in order to avoid false-positive
detections.

\begin{figure*}
  \resizebox{\hsize}{!}{\includegraphics{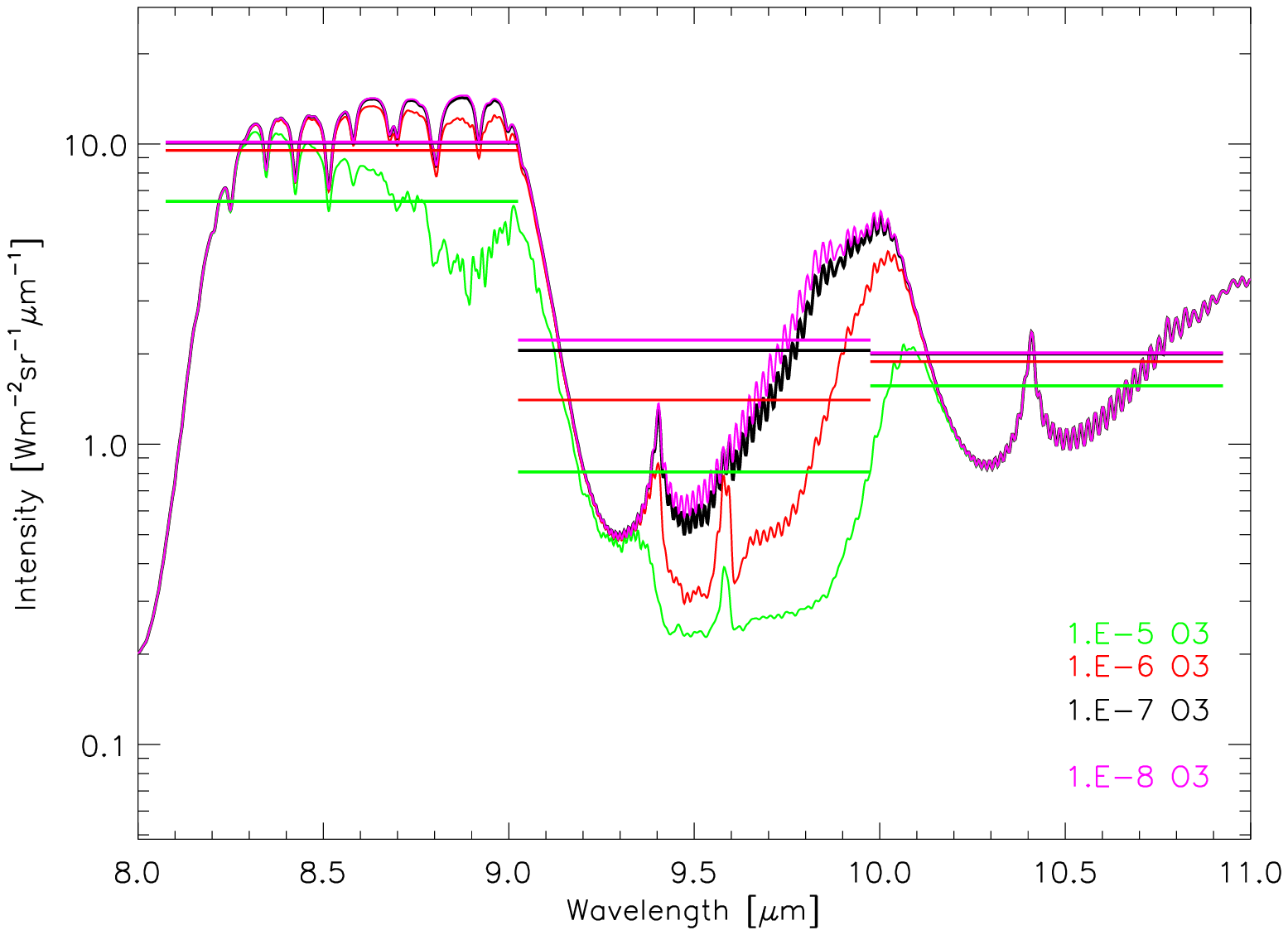}\\
  \includegraphics{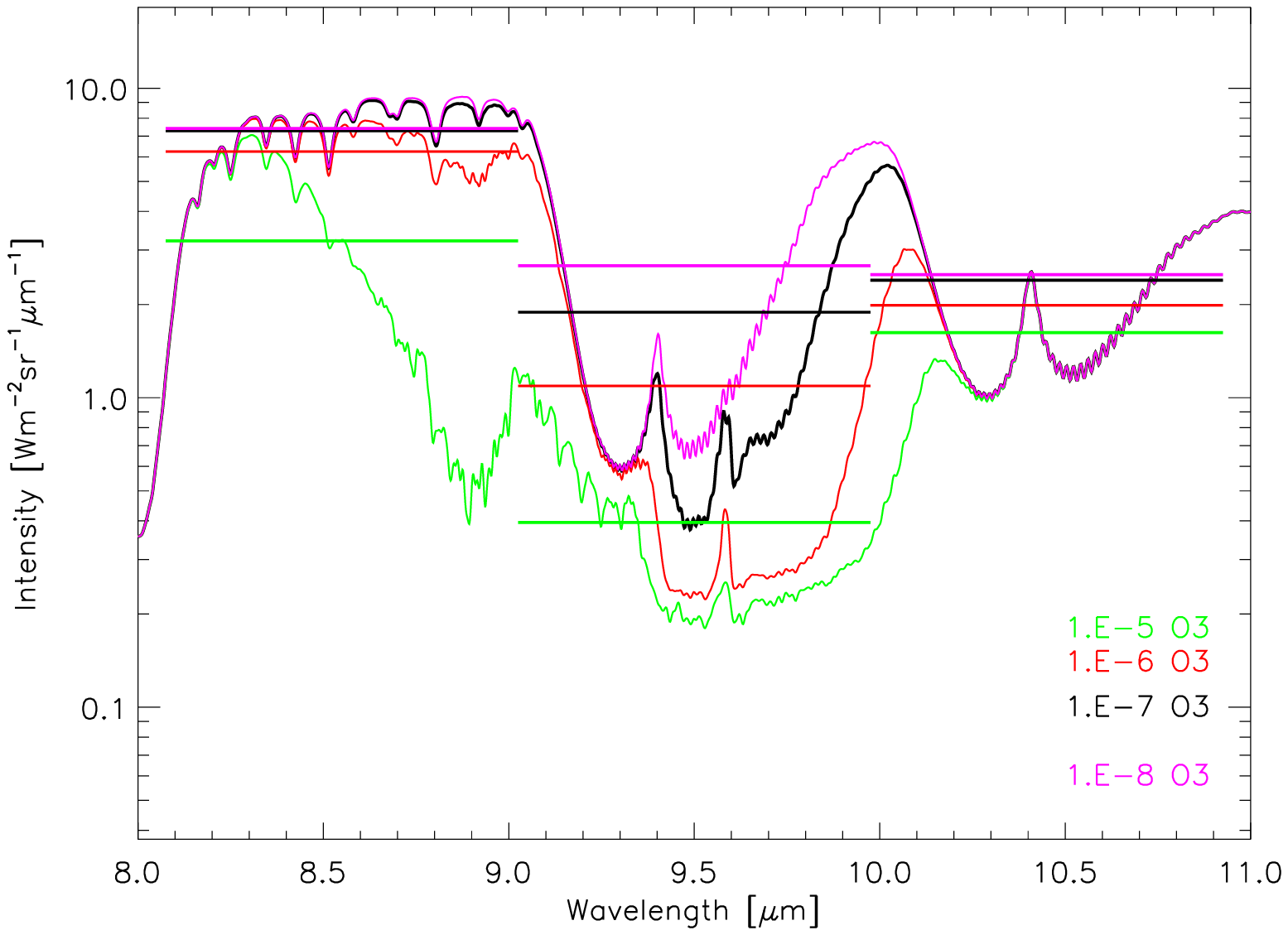}}\\
  \caption{Effect of O$_3$ isoprofiles on emission spectra. Constant O$_3$ concentrations as indicated. Left: High-CO$_2$ 20 bar, right: Medium-CO$_2$ 20 bar. Horizontal lines represent binned spectra for a spectral resolution of 10.}\label{falseneg}
\end{figure*}

\subsubsection{False-negative detection of biomarkers}

In order to investigate the possibility of false-negative detections of ozone, we calculated model spectra of the high-CO$_2$ 20 bar case and of the medium-CO$_2$ 20 bar case where we additionally introduced artificial profiles of ozone in the line-by-line spectral calculations. These ozone profiles have not been implemented in the 1D climate calculations. However, they are not expected to strongly influence the temperature structure of the atmosphere (e.g., \citealp{kaltenegger2011}) because of the lack of UV radiation emitted by GL 581 in the wavelength regime where ozone heating occurs. \newline In a first attempt, we used isoprofiles of ozone in the calculation of the spectra. Chosen concentrations were 10$^{-8}$, 10$^{-7}$, 10$^{-6}$ and 10$^{-5}$, respectively. This choice of concentrations broadly covers the range of ozone concentrations found in the present Earth atmosphere (10$^{-8}$ near the surface, about to almost 10$^{-5}$ in the mid-stratosphere).\newline Results of this sensitivity analysis are shown in Fig. \ref{falseneg} for the high-CO$_2$ and medium-CO$_2$ 20 bar cases. Clearly, a detection of ozone at low concentrations of 10$^{-8}$ or 10$^{-7}$ would not be possible at low spectral resolution. For the high-CO$_2$ 20 bar case, even a concentration of 10$^{-6}$ would be very challenging to detect. In both cases shown here, however, it would be possible to infer ozone levels of 10$^{-5}$ since the calculated intensity drops by about a factor of 3-8 in the center of the ozone fundamental band. 

In reality, ozone profiles will most likely not be in the form of isoprofiles. The production of ozone proceeds through the three-body reaction

\begin{equation}\label{o3prod}
 O_2+O+M\rightarrow O_3+M
\end{equation}

where $M$ is, on Earth, typically nitrogen. The production of the atomic oxygen needed in eq. \ref{o3prod} requires photolysis of molecular oxygen, carbon dioxide or water. Hence, there is a trade-off between the necessary UV radiation for photolysis and the density to enable the three-body production reaction, which is why photochemical models generally predict a distinct maximum of atmospheric ozone concentrations around the 1-10 millibar pressure range. \newline The actual location of this ozone maximum depends on the stellar UV radiation field, the ozone and the oxygen content of the atmosphere (\citealp{selsis2002}, \citealp{Seg2003}, \citealp{segura2007}, \citealp{domagal2010}, \citealp{grenfell2011}). Therefore, we inserted artificial ozone profiles $C_{O_3}$ as a function of pressure $p$ based on a Gaussian profile

\begin{equation}\label{o3profile_para}
 C_{O_3}\left(p\right)=C_{\rm{max}}\cdot exp\left(-\frac{\left(log\frac{p}{p_m}\right)^2}{0.5}\right)
\end{equation}

\begin{figure}[H]
  \includegraphics[width=220pt]{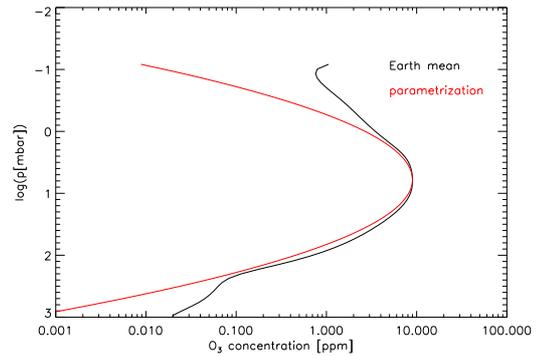}\\
  \caption{Modern Earth mean ozone profile from  \citet{grenfell2011} and approximation (in red) with eq. \ref{o3profile_para} using $C_{\rm{max}}$=9ppm and $p_m$=6 mbar. }\label{o3par}
\end{figure}

\begin{figure*}
  \resizebox{\hsize}{!}{\includegraphics{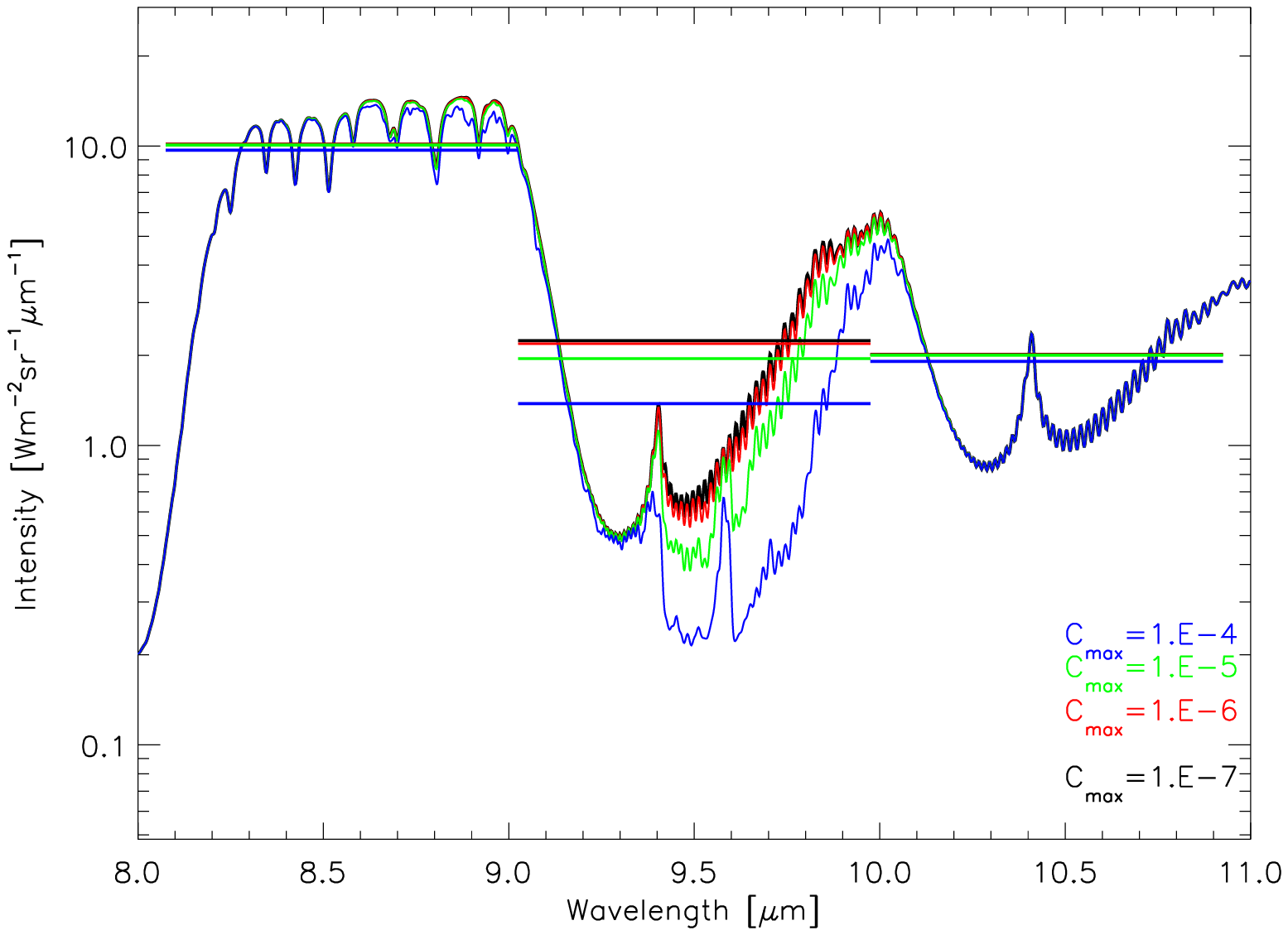}\\
  \includegraphics{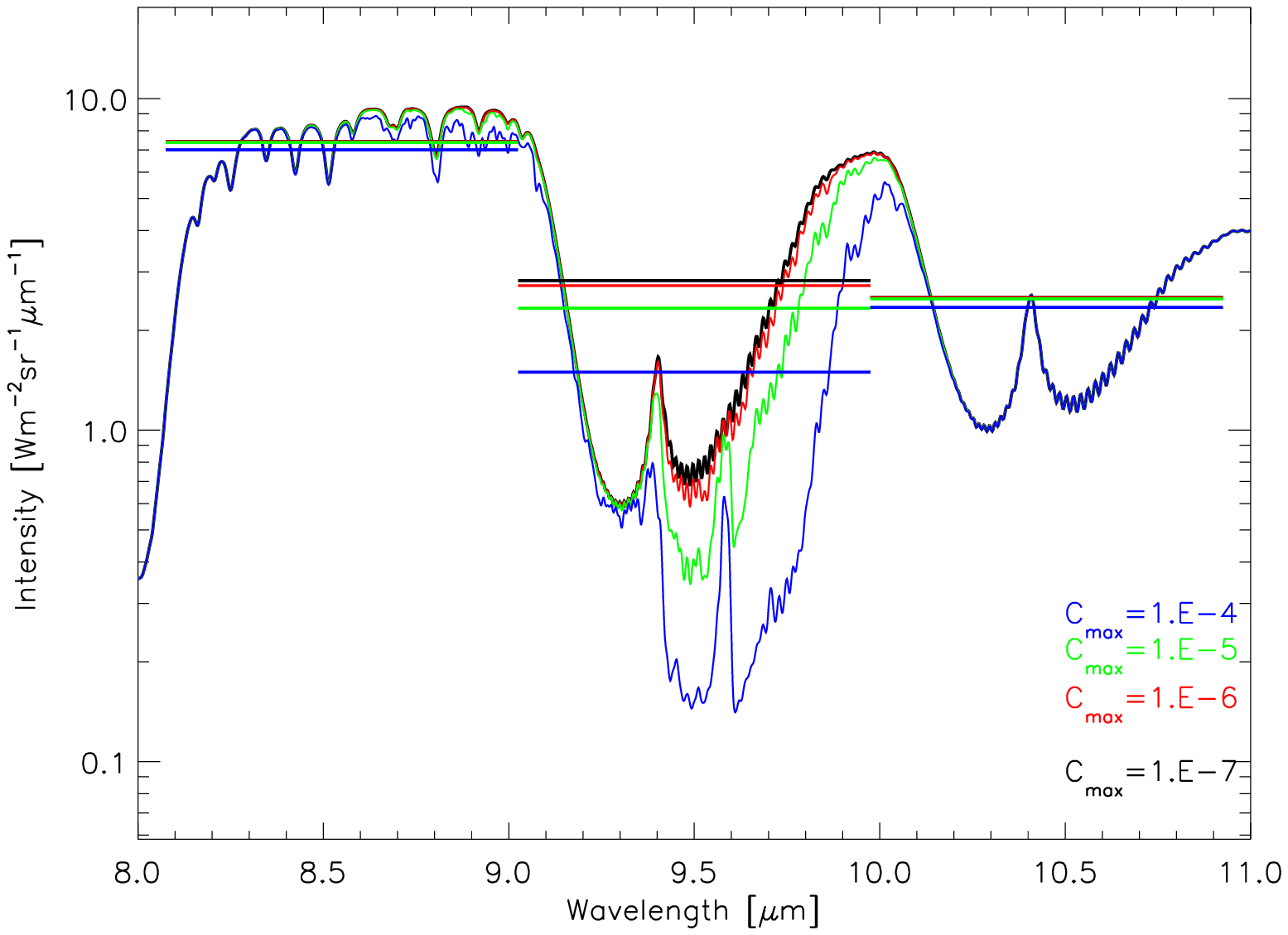}}\\
  \caption{Effect of O$_3$ profiles on emission spectra. Parameters for eq. \ref{o3profile_para}: $C_{\rm{max}}$ as indicated, $p_m$=10mbar. Left: High-CO$_2$ 20 bar, right: Medium-CO$_2$ 20 bar. Horizontal lines represent binned spectra for a spectral resolution of 10. }\label{falseneg_prof}
\end{figure*}
\begin{figure}[H]
\includegraphics[width=210pt]{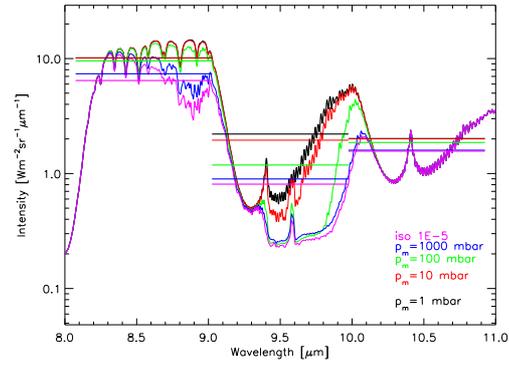}
\caption{Effect of maximum of O$_3$ profiles on emission spectra for the high-CO$_2$ 20 bar case.  Parameters for eq. \ref{o3profile_para}: $C_{\rm{max}}$=10ppm, $p_m$ in mbar as indicated. Horizontal lines represent binned spectra for a spectral resolution of 10.}
\label{falseneg_pmax}
\end{figure}


where $C_{\rm{max}}$ is the maximum ozone concentration reached at pressure $p_m$. Fig. \ref{o3par} shows the approximation of an Earth ozone profile with eq. \ref{o3profile_para}. The agreement near the maximum of the ozone layer is rather good. \newline Figure \ref{falseneg_prof} shows the effect of varying  $C_{\rm{max}}$ on emission spectra at constant $p_m$ of 10 mbar. Shown are the spectra for the high-CO$_2$ and medium-CO$_2$ 20 bar cases. In Fig. \ref{falseneg_pmax}, the effect of varying $p_m$ on the spectrum at constant  $C_{\rm{max}}$ of 10 ppm is illustrated for the high-CO$_2$ 20 bar case. \newline Note that tropospheric concentration maxima, i.e. at pressures higher than 100 mbar, are usually not expected from atmospheric chemistry modeling. Rather, maxima are found in the lower to upper stratosphere. Hence the scenarios with $p_m$ of 100 or 1,000 mbar shown in Fig. \ref{falseneg_pmax} are most likely overestimating the ozone column.\newline From Fig. \ref{falseneg_prof}, it is evident that it is very difficult to infer the presence of ozone even at high concentrations of about 10 ppm. Only for the profile with  $C_{\rm{max}}$=100ppm, it would be possible to detect an additional absorption besides the CO$_2$ band. \newline Figure \ref{falseneg_pmax} shows that for a value of $p_m$=100mbar or higher, it would be possible to detect ozone in the spectrum. However, as stated above, such a tropospheric maximum of ozone is not very probable.

These sensitivity studies illustrate that an inferred absence of ozone could be possible for the atmospheric scenarios considered in this work, even if ozone would be present in large concentrations of the order of 10$^{-6}$-10$^{-5}$. Hence, also the possibility of false-negatives of ozone needs to be taken into account when investigating CO$_2$-rich, potentially habitable atmospheric scenarios.

\subsection{Detectability}

\begin{table*}
  \centering
  \caption[SNR values for secondary eclipse spectroscopy]
  {SNR values for secondary eclipse spectroscopy. Telescope configuration JWST, integration time 4.15 hours, spectral resolution R=10.\newline}
  \label{emis_snr_jwst}
  \begin{tabular}{lccccc}
    \hline
     \hline
     Scenario             & 8.5 $\mu$m  & 9.5 $\mu$m  & 11.5 $\mu$m & 15 $\mu$m & 20 $\mu$m \\
     \hline
    \hline
     low-CO$_2$ 1 bar     & 0.050 & 0.091  & 0.19   &  0.10    & 0.23 \\
     low-CO$_2$ 20 bar    & 0.062 & 0.10    &  0.22 &  0.088   & 0.25 \\
     medium-CO$_2$ 1 bar  & 0.064 & 0.10   &   0.22 &  0.091   & 0.26\\
     medium-CO$_2$ 20 bar & 0.56 & 0.28   & 0.51   &  0.089   & 0.34 \\
     high-CO$_2$ 1 bar    & 0.12 & 0.15   & 0.37   &  0.10   & 0.34 \\
     high-CO$_2$ 20 bar   & 0.77 & 0.22   & 0.74   &  0.10   & 0.48 \\
      \end{tabular}
\end{table*}

\begin{table*}
  \centering
    \caption[SNR values for transmission spectroscopy]
  {SNR values for transmission spectroscopy. Telescope configuration JWST, integration time 4.15 hours, spectral resolution R=10.\newline}
  \label{transmis_snr_jwst}
  \begin{tabular}{lccccccc}
     \hline
    \hline
     Scenario              &  2.0 $\mu$m & 2.7 $\mu$m & 4.3 $\mu$m & 6.3 $\mu$m & 7.7 $\mu$m & 9.5 $\mu$m     & 15 $\mu$m \\
     \hline
    \hline
     low-CO$_2$ 1 bar      &  0.41       & 0.65       & 0.83       & 0.091      & 0.027      & 8.4 10$^{-3}$  & 0.21 \\
     low-CO$_2$ 20 bar     &  1.1        & 1.2        & 1.4        & 0.24       & 0.22       & 0.12           & 0.33 \\
     medium-CO$_2$ 1 bar   &  1.0        & 1.1        & 1.2        & 0.12       & 0.23       & 0.12           & 0.29 \\
     medium-CO$_2$ 20 bar  &  2.1        & 1.9        & 2.0        & 0.46       & 0.65       & 0.46           & 0.44 \\
     high-CO$_2$ 1 bar     &  0.96       & 0.89       & 0.94       & 0.10       & 0.26       & 0.15           & 0.22 \\
     high-CO$_2$ 20 bar    &  1.9        & 1.6        & 1.6        & 0.50       & 0.65      & 0.46           & 0.35 \\
  \end{tabular}
\end{table*}

The main spectral bands investigated for detectability are the 2.0,
2.7, 4.3, 7.7, 9.5 and 15 $\mu$m CO$_2$ absorption bands as well as
the 6.3 $\mu$m H$_2$O absorption band in transmission. For secondary eclipse spectroscopy, only the 9.5 and 15\,$\mu$m CO$_2$ and the 20 $\mu$m H$_2$O bands were considered. Additionally two filters outside of broad absorption bands are calculated, namely at 8.5 and at 11.5 $\mu$m. These two filters offer the possibility to characterize surface temperatures in optically thin atmospheres.  \newline Furthermore, six
re\-presentative scenarios are considered, namely the 1 and 20 bar
runs of the low, medium and high-CO$_2$ cases.

Table \ref{emis_snr_jwst} summarizes the SNR values for
secondary eclipse spectroscopy. It is clearly seen that,
even though GL 581 is a very close star (6.27 pc,
\citealp{butler2006}), obtainable SNR values are very small,
only reaching values of 0.3-0.77. The reason for the low SNR values in the mid- to far IR is the effect
of the zodiacal background chosen here (see Appendix
\ref{appendix_noise}) which reduces SNR values by about a factor of
5 at 20 $\mu$m and a factor of 2 at 15 $\mu$m compared to the
photon-limited case. A similar result was already found by, e.g.,
\citet{belu2010}.

Table \ref{transmis_snr_jwst} summarizes the SNR values for
transmission spectroscopy. SNR values are mostly below unity. The
exceptions are the 2.0, 2.7 and 4.3 $\mu$m near-IR fundamentals of
CO$_2$. In these bands, characterization of the GL\,581\,d
scenarios, as outlined above, could be feasible. A clear
distinction between, e.g., high and low-CO$_2$ cases is still very
difficult, given that the SNR values are only marginally larger than
unity. One strategy to obtain larger SNRs is the co-adding of transits. This might result in a significant increase of the SNR since the SNR scales with the square root of the number of transits. Assuming a 5-year mission lifetime of JWST (3-4 transits observable per year) would lead to SNRs of about a factor of 4 higher compared to the values shown in Tables \ref{emis_snr_jwst} and \ref{transmis_snr_jwst}.

\section{Conclusions}

\label{concl}

We presented synthetic emission and transmission spectra for a wide range of atmospheric scenarios  of the potentially habitable
Super-Earth GL\,581\,d. These spectra were used as an example of possible
future spectroscopic investigations of candidate habitable planets.

Water and carbon dioxide could be clearly seen in the calculated
spectra due to prominent absorption bands, indicating the presence
of an atmosphere.

The determination of surface temperatures was possible for model atmospheres with either low surface pressures or low CO$_2$ content. The potentially habitable, CO$_2$-rich scenarios did not allow for the characterization of surface temperatures. Thus, their potential habitability could not be assessed directly from the spectra, in agreement with a model spectrum from \citet{kaltenegger2011}.  The further determination of atmospheric conditions was complicated by degeneracies between the surface pressure and the CO$_2$ concentration. However, when combining observations in several spectral bands and using both transmission and emission spectra, inferring approximative CO$_2$ concentrations and surface pressures would be possible.
  
With currently planned
telescope designs such as JWST, SNR values for
emission and transmission spectroscopy are, however, rather low,
implying that the detection of an atmosphere of a "GL\,581\,d-like"
transiting planet is challenging. Reasonable, single-transit SNR
could only be calculated for three near-IR CO$_2$ fundamentals at
2.0, 2.7 and 4.3 $\mu$m (transmission spectroscopy). In the mid- to far-IR, thermal and zodiacal background noises inhibit SNRs above unity.

Results indicate that the search for biomarkers in CO$_2$-rich or
high-pressure atmospheres would suffer from the possibility of
false-positive detections, in agreement with previous studies (e.g., \citealp{schindler2000} or \citealp{selsis2002}). This is due to absorption bands of CO$_2$
which occur close to main biomarker absorption bands. However, if
the main CO$_2$ IR bands were to be observed simultaneously, such
false-positive detections could possibly be avoided. This will be
subject of future modeling including Earth-like atmospheric
composition. Results also imply that CO$_2$ absorption bands could mask the spectroscopic features of ozone, hence produce false-negative detections.  This was shown to be possible even if ozone would be present in rather large concentrations of up to 10$^{-5}$.

\begin{acknowledgements}

This research has been supported by the Helmholtz Association
through the research alliance "Planetary Evolution and Life". Philip von Paris and Pascal Hedelt acknowledge support from the European Research
Council (Starting Grant 209622: E$_3$ARTHs). Insightful discussions with A.B.C. Patzer and F. Selsis are gratefully acknowledged. We thank the anonymous referee for his/her constructive
remarks which helped to improve the paper.

\end{acknowledgements}

\begin{appendix}

\section{Noise contributions}
\label{appendix_noise}

The total SNR $S_p$ of a planetary feature is calculated
with

\begin{equation}\label{planetsnr}
 S_p=\frac{F_p}{\sigma_{\rm{total}}}
\end{equation}

where $F_p$ is the planetary signal measured during
secondary eclipse (emitted and reflected spectrum) or primary eclipse
(additional transit depth due to atmospheric absorption).
$\sigma_{\rm{total}}$ represents the total noise of the measurement.
In addition to the stellar noise $\sigma_S$, main contributions to
the noise budget come from the zodiacal emission of the solar system
(denoted $\sigma_{\rm{zodi}}$), the thermal emission of the
mirror and, in the case of the JWST, the sun shade (denoted
$\sigma_{\rm{thermal}}$). Both components need to be taken into
account in the SNR calculations since their effect on detectability
could be significant, especially in the mid-IR, longwards of 10
$\mu$m (e.g., \citealp{deming2009}, \citealp{belu2010}). In addition, the dark noise $\sigma_d$ is considered here. Hence, we
have

\begin{equation}\label{totalnoise}
    \sigma_{\rm{total}}=\sqrt{2\cdot(\sigma_S^2+\sigma^2_{\rm{thermal}}+\sigma^2_{\rm{zodi}}+\sigma_d^2)}
\end{equation}

The thermal noise was obtained from

\begin{equation}\label{thermalnoise}
 \sigma_{\rm{thermal}}^2=q_e \cdot \epsilon \cdot B_{\lambda}(T_{\rm{JWST}})\cdot
 t_t \cdot  n_{\rm{px}} \cdot S_A \cdot A_E \cdot \frac{\lambda}{R} 
\end{equation}

where $\epsilon$ is the emissivity, $B_{\lambda}(T_{\rm{JWST}})$
the blackbody emission of the telescope and sun shade, $t_t$ the integration time
(here assumed to be the transit duration, i.e. 4.15 hours),
$n_{\rm{px}}$ the number of pixels used for the integration on the detector, $\lambda$ the considered wavelength and $R$ the spectral resolution, $A_E$ the emitting area, $q_e$=0.15 the quantum efficiency of the telescope
(e.g., \citealp{Kaltenegger2009}, \citealp{rauer2011}) and $S_A$ the solid angle per pixel.\newline We assumed a telescope
and sun shade temperature $T_{\rm{JWST}}$ of 45 K with an emissivity
of $\epsilon$=0.15 \citep{belu2010} and a total emitting surface of
sun shade and mirror combined of $A_E$=240 m$^2$ \citep{nella2002}. The number of pixels $n_{\rm{px}}$ is
calculated with 

\begin{equation}\label{numberpix}
    n_{\rm{px}}= n_{\rm{spatial}}\cdot n_{\rm{spectral}} \cdot \frac{\lambda}{R}
\end{equation}

where $n_{\rm{spatial}}$ and $n_{\rm{spectral}}$ the number of pixels in
spatial and spectral (per $\mu$m) direction, respectively. We assumed $n_{\rm{spectral}}$=32.33 pixel $\mu$m$^{-1}$, mimicking the MIRI
instrument in the 5-11 $\mu$m spectral range \citep{belu2010}.
 The number of pixels in the spatial direction is obtained from 
 
\begin{equation}\label{nspat}
    n_{\rm{spatial}}=2.4\cdot \frac{\lambda}{d_t\cdot p_s}
\end{equation}

with $d_t$ the telescope diameter (i.e., 6.5 m in the case of JWST) and $p_s$=0.11" per pixel the instrumental pixel scale. The number calculated from eq. \ref{nspat} is then rounded up to the closest multiple of 4.  The solid angle per pixel is calculated with (1
sq. deg $\approx \frac{1}{3282}$ sr)

\begin{equation}\label{solidangle}
    S_A=\frac{p_s^2}{3600^2 \cdot 3282}
\end{equation}

The zodiacal emission $I_Z$ (given in W m$^{-2}$
$\mu$m$^{-1}$ sr$^{-1}$) used in the noise calculations is taken
from one example measurement in the ecliptic plane presented by
\citet{kelsall1998} (their Fig. 9). Note that this choice is a rather pessimistic assumption since some potential targets will probably be located towards higher ecliptic latitudes, hence the corresponding zodiacal noise would be somewhat reduced. We calculate the zodiacal noise via

\begin{equation}\label{zodinoise}
\sigma_{\rm{zodi}}^2=I_Z\cdot
n_{\rm{px}} \cdot S_A \cdot A_{\rm{tel}} \cdot
t_t \cdot q_e \cdot \frac{\lambda}{R}
\end{equation}

with $A_{\rm{tel}}$ the telescope area, assuming a circular aperture of 6.5 m.\newline The dark noise  contribution to the noise budget is
calculated from

\begin{equation}\label{dark}
    \sigma_d^2=d_c \cdot n_{\rm{px}} \cdot t_t
\end{equation}

where $d_c$=0.03 e$^-$ pixel$^{-1}$ s$^{-1}$ is the dark current (MIRI, see e.g. \citealp{deming2009} or \citealp{belu2010}).

\end{appendix}

\bibliographystyle{aa}
\bibliography{literatur_glspec}

\end{document}